\PassOptionsToPackage{table}{xcolor}
\documentclass[%
 reprint,
superscriptaddress,
footinbib,
 amsmath,amssymb,
 aps,
pra,
]{revtex4-2}

\usepackage{tikz} 
\usepackage{amsmath,amsfonts,amsthm, mathtools}
\usepackage{float}
\usepackage{dsfont}
\usepackage{csquotes}
\usepackage{amssymb}
\usepackage{braket}
\usepackage[table]{xcolor}
\usepackage{verbatim}
\usepackage{mathtools}
\usepackage{rotating}
\usepackage{thmtools, thm-restate}
\usepackage{bm}
\usepackage{hyperref}
\usepackage{mathtools}
\usepackage{mathbbol}
\usepackage{makecell}
\usepackage{pifont}
\usepackage{appendix}
\usepackage{diagbox}      
\usepackage{soul}

\definecolor{darkred}  {rgb}{0.5,0,0}
\definecolor{darkblue} {rgb}{0,0,0.5}
\definecolor{darkgreen}{rgb}{0,0.5,0}
\hypersetup{
  colorlinks = true,
  urlcolor  = blue,         
  linkcolor = darkblue,     
  citecolor = darkgreen,    
  filecolor = darkred    
}
\theoremstyle{definition}

\newtheorem{lemma}{Lemma}
\newtheorem{proposition}{Proposition}

\newtheorem{theorem}{Theorem}

\newtheorem*{remark}{Remark}

\newcommand{\mbf}{\mathbf}
\newcommand{\mbb}{\mathbb}
\newcommand{\mc}{\mathcal}

\newcommand{\tr}{\textrm{Tr}}

\usepackage{multirow}
\newcommand{\op}[2]{|#1\rangle\langle#2|}

\definecolor{cool_green}{rgb}{0.0, 0.5, 0.0}

\usepackage{blkarray}
\usepackage{adjustbox}
\usepackage{graphicx}
\usepackage{dcolumn}
\usepackage{bm}

\usepackage[normalem]{ulem}

\newcommand{\yujie}{\color{black}}
\newcommand{\blk}{\color{black}}

\newcommand{\yk}{\color{black}}

\begin{document}

\preprint{APS/123-QED}


\title{Quantum-Limited Subdiffraction Telescopy Requires Genuine Multi-Telescope Interference}


\author{Yujie Zhang}
\email{yujie4physics@gmail.com}
\affiliation{Institute for Quantum Computing and Department of Physics \& Astronomy,
University of Waterloo, 200 University Ave W, Waterloo, Ontario, N2L 3G1, Canada}
\affiliation{Perimeter Institute for Theoretical Physics, 31 Caroline Street North, Waterloo, Ontario Canada N2L 2Y5}

\author{Yunkai Wang}
\affiliation{Perimeter Institute for Theoretical Physics, 31 Caroline Street North, Waterloo, Ontario Canada N2L 2Y5}
\affiliation{Institute for Quantum Computing and Department of Applied Mathematics,
University of Waterloo, 200 University Ave W, Waterloo, Ontario, N2L 3G1, Canada}

\author{Wilson Wu}
\affiliation{Department of Physics, Simon Fraser University, 8888 University Dr W, Burnaby, BC V5A 1S6, Canada}

\author{Thomas Jennewein}
\affiliation{Institute for Quantum Computing and Department of Physics \& Astronomy,
University of Waterloo, 200 University Ave W, Waterloo, Ontario, N2L 3G1, Canada}
\affiliation{Department of Physics, Simon Fraser University, 8888 University Dr W, Burnaby, BC V5A 1S6, Canada}
\date{\today}

\begin{abstract}
\yujie 
Conventional stellar interferometry reconstructs incoherent sources from pairwise mutual coherences between telescopes. Are such pairwise measurements sufficient for quantum-limited subdiffraction imaging with a telescope array? We show that for generic image-moment estimation, they are not. We consider weak incoherent light from a generic extended source observed by an array of telescopes, each supporting a single optical mode. For an $N$-telescope array, we derive the quantum Fisher information (QFI) scaling of image moments up to the cutoff $2N-2$ and prove that arbitrary measurements restricted to telescope pairs attain the full-array QFI scaling only up to second order. Thus, estimating higher-order moments at the quantum limit requires genuinely multi-telescope interference. Inspired by spatial-mode demultiplexing (SPADE) from single-aperture subdiffraction imaging, we construct array-SPADE measurements that attain the optimal QFI scaling up to the finite-array cutoff. Finally, we show that these measurements can, in principle, be embedded in ancilla- and memory-assisted quantum-network architectures for long-baseline telescopy. \blk
\end{abstract}

\maketitle
The angular resolution of a conventional imaging system is commonly associated with the diffraction scale $\theta_D\sim1/(kD)$, where $D$ is the effective aperture size and $k$ is the optical wavenumber. However, the quantum theory of superresolution has shown that this diffraction scale is not a fundamental limit for estimating parameters of incoherent sources~\cite{Tsang2016,Tsang2019Review}. This has led to spatial-mode demultiplexing (SPADE) and related measurement procedures that overcome the direct-imaging limits~\cite{Paur16,Lupo2016,Nair2016,Ang2017,Tham2017,Tsang2017,Tsang2018,Tsang2018a,Wang21,Ou2021,Zanforlin2022}. The framework has also been extended beyond point-source models to the estimation of general extended incoherent sources in the subdiffraction regime. In this regime, low-order moments characterize coarse properties such as centroid and extent, while higher-order moments capture progressively finer source features~\cite{Tsang2019,Tsang2019Semiparametric,ZhouJiang2019,Dutton2019,Tsang2021,Sorelli2021a,Sorelli2021b,Tan2023,Wang2023, Olmi2026,Wang2025d}.

Long-baseline telescope arrays provide a complementary route to high angular resolution by synthesizing a large effective aperture from separated telescopes. In conventional stellar interferometry, the van Cittert--Zernike theorem relates the source intensity distribution to pairwise mutual coherences between telescopes, so standard aperture synthesis is organized baseline by baseline~\cite{Cittert1934,Zernike1938,Goodman2017,Monnier2003}. Recent work has begun to extend quantum-superresolution ideas to long-baseline telescope-array architectures~\cite{lupo2020,wang2021,Hu2025,Sajjad2024,Padilla2026,Padilla2026a}. Building on the analysis of Ref.~\cite{Sajjad2024}, Padilla et al.~\cite{Padilla2026,Padilla2026a} show that telescope arrays, when supplemented by local spatial-mode sorting and entanglement-assisted inter-telescope processing, can attain quantum limits for important point-source models. Their results clarify the complementary roles of intra-telescope spatial-mode information and inter-telescope baseline information. However, the explicit analytical examples developed so far mainly concern low-dimensional point-source models, such as two-point sources, where the local estimation problem is governed by low-order moment information and pairwise interference can already be sufficient.  This leaves the central question open: are arbitrary pairwise interferometric measurements sufficient to attain the full-array quantum limit for subdiffraction image-moment estimation, or is genuinely multi-telescope interference necessary?

To address this question, we consider general incoherent extended sources, described by their normalized image moments, and isolate the interferometric array itself by working in a single-mode-per-telescope model. This removes intra-telescope spatial structure, so any remaining advantage must come from coherent processing of multiple telescope modes. We derive the QFI scaling of normalized image moments, summarized in Table~\ref{tab:main}, and prove that pairwise measurement schemes attain the optimal full-array scaling only for the lowest-order moments. For moments $3\le \mu\le 2N-2$, reaching the full-array quantum limit requires genuine multi-telescope interference, i.e., one-photon measurement effects with coherent support on three or more telescope modes.

We then construct array-SPADE measurements that attain the full-array QFI scaling. These measurements can be viewed as a synthesized-aperture analogue of SPADE used in single-aperture quantum superresolution. Furthermore, recent work on quantum-assisted telescopy has shown that shared entanglement-assisted resources and quantum memories can implement nonlocal interferometric measurements over long baselines~\cite{Gottesman2012,tsang2011,Khabiboulline2019PRA,Khabiboulline2019PRL,Huang2022,wang2023astronomical,huang2024,Wang2025,Czupryniak2023,Marchese2023,Zhang2025, Wang2025b, Huang2026}, with recent experimental demonstrations of the relevant nonlocal interferometry~\cite{Matthew2023,Diaz21,Stas2026,Wang2026}. We also show that the array-SPADE measurements identified here can, in principle, be embedded into ancilla- and memory-assisted quantum-telescope architectures~\cite{Gottesman2012,Khabiboulline2019PRL,Stas2026}. 

\begin{table}[h]
\centering
\caption{
QFI $\mbb{H}_{\mu\mu}$ or FI $\mbb{F}_{\mu\mu}$ for estimating the normalized image moment $x_\mu$ ($\mu\ge 1$) in the weak-light, subdiffraction regime $(\epsilon,\Delta\ll1)$, where $\epsilon$ is the mean photon number per temporal mode and $\Delta$ is the subdiffraction parameter. The table gives the
leading asymptotic scaling; $O$ and $\Theta$ denote upper bounds and matching upper/lower bounds, respectively. Except for the single-aperture scheme, the listed schemes use no intra-telescope spatial-mode structure.  Here $\lceil x \rceil$ denotes the ceiling of $x$, i.e., the smallest integer no less than $x$.
}
\begin{ruledtabular}
\begin{tabular}{l l l}
\label{tab:main}
Scheme  & \multicolumn{2}{l}{$\mbb{H}_{\mu\mu}(\epsilon,\Delta)$ or $\mbb{F}_{\mu\mu}(\epsilon,\Delta)$} \\
\hline
Single aperture~\cite{Tsang2019,ZhouJiang2019}
& $\Theta\bigl(\epsilon\Delta^{2\lceil \mu/2\rceil}\bigr)$
& all $\mu$ \\
\hline

\multirow{2}{*}{Full $N$-telescope array (this work)}
& $\Theta\bigl(\epsilon\Delta^{2\lceil \mu/2\rceil}\bigr)$
& $\mu \le 2N-2,$ \\

& $O\bigl(\epsilon\Delta^{2\mu-2N+2}\bigr)$
& $\mu > 2N-2$ \\
\hline

\multirow{2}{*}{Pairwise optimum}

& $\Theta\bigl(\epsilon\Delta^{2\mu-2}\bigr)$
& $\mu ~\text{even}$, \\

& $\Theta\bigl(\epsilon\Delta^{2\mu}\bigr)$
& $\mu ~\text{odd}$. \\

\hline{Pairwise heterodyne/homodyne}
& $\Theta\bigl(\epsilon^2\Delta^{2\mu}\bigr)$
& all $\mu$ \\

\hline
\multirow{2}{*}{Pairwise intensity interferometry}
& $\Theta\bigl(\epsilon^2\Delta^{2\mu}\bigr)$
& $\mu~\text{even},$ \\
& $O\bigl(\epsilon^2\Delta^{2\mu+2}\bigr)$
& $\mu~\text{odd}$ \\
\end{tabular}
\end{ruledtabular}
\end{table}

\yk The full-array and pairwise QFI scalings in Table~\ref{tab:main} are derived below, while the FI scalings for specific pairwise schemes are discussed in the Supplemental Material~\cite{SM}. The dimensionless subdiffraction parameter $\Delta$ measures the source size $L$ relative to the source-plane diffraction length $z_0\theta_D$, with $\theta_D=1/(kD)$, i.e., 
\begin{align}
\Delta:=\frac{kLD}{z_0}.
\label{eq:delta-def}
\end{align}

For a single aperture, $D$ is the aperture size; for a telescope array, $D$ is the largest baseline.  Because $D$ is the largest baseline for an array, matching the single-aperture $\Delta$-scaling can still correspond to much higher angular sensitivity~\footnote{For example, in estimating the source separation of a balanced two-point-source model, the QFI is dominated by the interferometric contribution rather than by intra-telescope mode sorting, e.g., Eq.~(94) of Ref.~\cite{Sajjad2024}.}. 
\blk 
\begin{figure}
    \centering
    \includegraphics[width=0.95\linewidth]{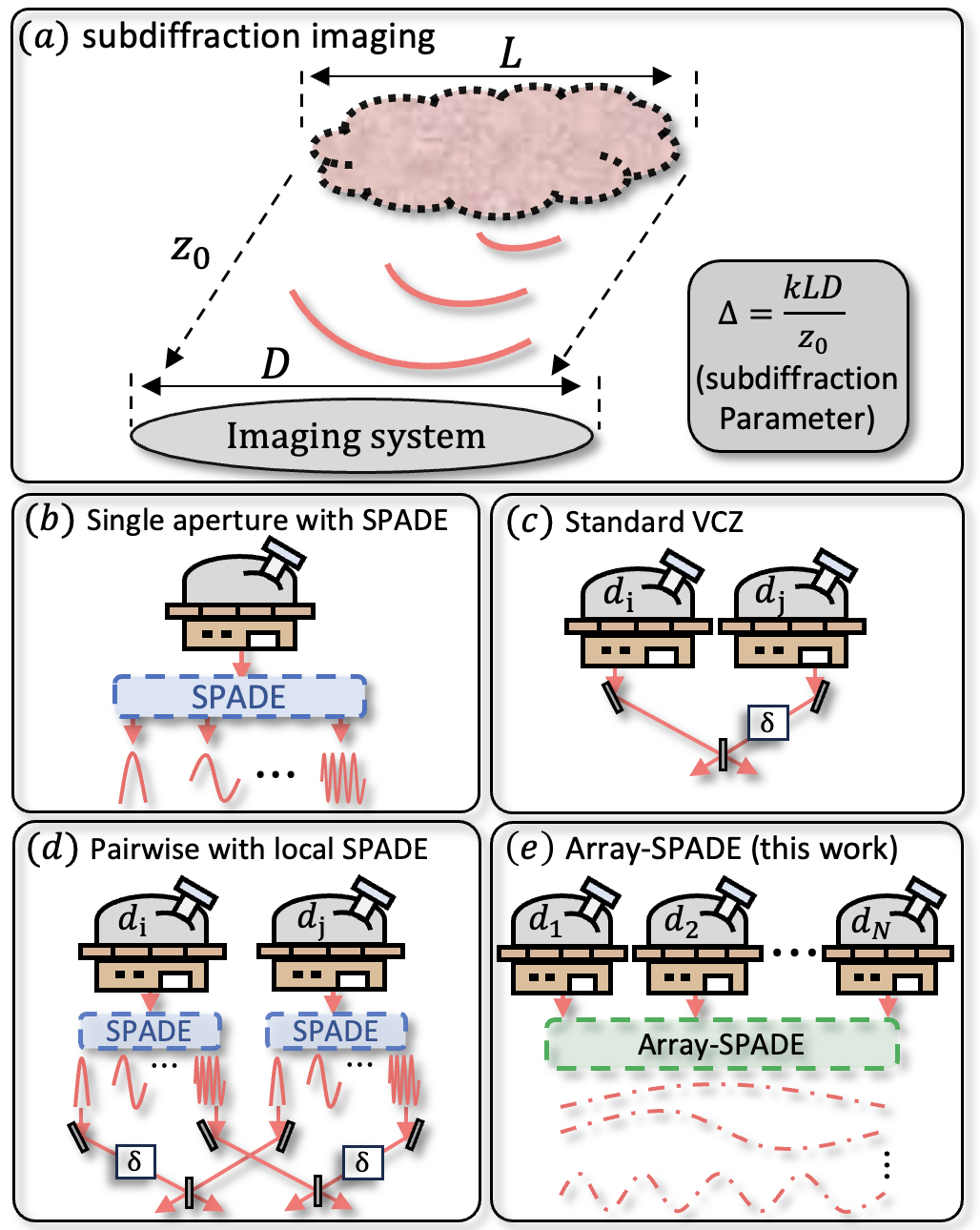}
\caption{Measurement architectures for subdiffraction imaging.
(a) Geometry of an extended incoherent source in the subdiffraction regime.
(b) Single-aperture SPADE sorts intra-aperture spatial modes~\cite{Tsang2016}.
(c) Standard VCZ interferometry estimates pairwise complex visibilities baseline by baseline~\cite{Monnier2003}.
(d) Pairwise schemes with local SPADE first sort spatial modes at each telescope and then interfere selected modes pairwise~\cite{Padilla2026,Padilla2026a}.
(e) This work: Array-SPADE implements full-array joint projections across the telescope modes, without using intra-telescope spatial-mode sorting.
}\label{fig:interferometry}
\end{figure}
\\
\textit{Preliminaries.--}
Consider an $N$-telescope array in one dimension with telescope coordinates
$d_j$. We define
\begin{align}
D:=\max_{j,k}|d_j-d_k|,
\quad
c_j:=\frac{d_j}{D},
\quad
c_{jk}:=c_j-c_k ,
\end{align}
so that $D$ is the largest physical baseline and $\{c_j\}$ specifies the
dimensionless array geometry. 


In this subdiffraction regime with $\Delta\ll1$ , it is useful to describe the source by
its normalized image moments about a reference point $u_0$,
\begin{align}
x_\mu:=\int du I(u)\left(\frac{u-u_0}{L}\right)^\mu,
\label{eq:image-moments}
\end{align}
where $u$ is the transverse coordinate in the source plane, $I(u)$ is the normalized source intensity distribution, and therefore $x_0=1$. Throughout the analysis, we treat $u_0$ as fixed, either known a priori or obtained from a preliminary localization step, and the moments $\{x_\mu\}$ are also treated as dimensionless source-shape parameters and are thus independent of $\Delta$.

To connect this moment description with conventional aperture synthesis, recall that the van Cittert--Zernike theorem (VCZ) relates pairwise mutual coherence to Fourier components of the source intensity distribution~\cite{Cittert1934,Zernike1938}. For a one-dimensional incoherent source, the complex visibility between telescopes $j$ and $k$ is
\begin{align}
g_{jk}&=\int duI(u)\exp\left[-i k\left(\frac{u(d_j-d_k)}{z_0}\right)\right] \notag \\
&=e^{i\phi_{jk}}
\sum_{\mu\ge0}
\frac{(-ic_{jk}\Delta)^\mu}{\mu!} x_\mu 
\label{eq:gij}
\end{align}
where $\phi_{jk}=\phi_j-\phi_k$ with
$\phi_j=-ku_0d_j/z_0$. \yk The VCZ theorem allows one to use the mutual coherence $\{g_{jk}\}$ as the primary observables; here we instead treat the moments $\{x_\mu\}$ as the parameters to be estimated.\blk

In optical stellar interferometry, the relevant signal typically lies in
the weak-light regime, with mean photon number per temporal mode
$\epsilon\ll1$~\cite{tsang2011}. It is therefore standard to expand the signal state in photon number as
\begin{align}
\rho_s=(1-\epsilon)\rho_s^{(0)}+
\epsilon\rho_s^{(1)}+O(\epsilon^2),
\label{eq:rho}
\end{align}
where $\rho_s^{(0)}$ and $\rho_s^{(1)}$ are the vacuum and one-photon
components. To leading order in $\epsilon$, all interferometric information is contained in the one-photon component. 

In the single-photon, $N$-mode Fock basis $\{\ket{1_j}_s\}_{j=1}^N$ with $\ket{1_j}_s$ denoting the state with one photon in mode $s_j$, the one-photon state $\rho_s^{(1)}$ takes the form
\begin{align}
\rho^{(1)}_s=\frac{1}{N}\sum_{j,k=1}^{N} g_{jk}\op{1_j}{1_k}_s
\label{eq:one-photon-rho}
\end{align}

Expanding $g_{jk}$ in powers of $\Delta$ as in Eq.~\eqref{eq:gij} gives the moment
decomposition of the one-photon state,
\begin{align}
\rho_s^{(1)}
&=
\sum_{\mu\ge0}x_\mu A_\mu,
\nonumber\\
A_\mu
&=
\Delta^\mu
\sum_{m+n=\mu}
\frac{i^{n-m}}{m!n!}
H^m\ket{\psi}\bra{\psi}H^n ,
\label{eq:A-mu-prelim}
\end{align}
where
$\ket{\psi}=N^{-1/2}\sum_j e^{i\phi_j}\ket{1_j}_s$ is the point-source state at $u_0$, and $H=\mathrm{diag}(c_1,\ldots,c_N)$ encodes the dimensionless array geometry. Thus $x_\mu$ enters only at order $\Delta^\mu$.


\yujie

\blk


\textit{QFI scaling--} We use the quantum Fisher information (QFI) to quantify the ultimate precision for estimating the moment parameter $x_\mu$. Let
\begin{align}
\rho^{(1)}_s=\sum_{q=0}^{N-1}\lambda_q\ket{q}\bra{q}
\label{eq:eigen}
\end{align}
be the spectral decomposition of the one-photon source state. Since $\partial_{x_\mu}\rho^{(1)}_s=A_\mu$ per Eq.~\eqref{eq:A-mu-prelim}, the QFI for estimating $x_\mu$ is given by
\begin{align}
\mbb H_{\mu\mu}(\rho_s^{(1)})
=\sum_{p,q:\lambda_p+\lambda_q>0}
\frac{2\left|\bra{p}A_\mu\ket{q}\right|^2}{\lambda_p+\lambda_q}.
\label{eq:QFI}
\end{align}


To extract the subdiffraction QFI scaling, it is convenient to introduce the subspace 
\begin{align}
\mc K:=\mathrm{Span}\{\ket{\psi},H\ket{\psi},H^2\ket{\psi},\dots\}.
\end{align}

On the telescope array side, distinct normalized coordinates $c_j$ imply $\dim \mc K=N$, and the Gram--Schmidt orthogonalization of $\mc K$ yields an orthonormal basis $\{\ket{e_a}\}_{a=0}^{N-1}$ satisfying
\begin{align}
\bra{e_a}H^m\ket{\psi}=0, \qquad m<a, \notag\\
\bra{e_a}H^m\ket{\psi}\neq 0, \qquad m= a.
\label{GSbasis}
\end{align}
This basis depends only on the array geometry $\{c_j\}$ and the reference point $u_0$; it does not depend on the unknown source moments. We call it the array-SPADE basis because it can be viewed as the synthesized-aperture counterpart of the basis of the single-aperture SPADE~\cite{Tsang2016}.
A more explicit construction, including its relation to single-aperture SPADE through a finite spatial Fourier transform, is given in the Supplemental Material~\cite{SM}. 
\blk 

On the source side, we assume a generic \textit{nondegeneracy} condition,  analogous to the one made in Ref.~\cite{Tsang2019}, that the $N\times N$ normalized Hankel matrix $\Gamma$ with
\begin{align}
\Gamma_{q,p}:=x_{p+q},\quad q,p \in \{0,\cdots N-1\}
\end{align}
is positive definite, i.e., $\Gamma>0$. Physically, this excludes source families with rank-deficient moment structure, such as discrete sources supported on fewer than $N$ points, but is otherwise generic.

Appendix~\ref{app:proofs} shows that the
ordered eigenvalues of $\rho_s^{(1)}$ satisfy
$\lambda_q=\Theta(\Delta^{2q})$, while the eigenvectors are localized
near the corresponding array-SPADE modes, i.e.,  $|\braket{e_a|q}|=O(\Delta^{|a-q|})$ for $a\ne q$ and $|\braket{e_q|q}|=\Theta(1)$.

These estimates yield the following local single-parameter QFI scaling for estimating image moment $x_{\mu}$.
\begin{theorem}
\label{Thm:qfi-scaling}
Under the \textit{nondegeneracy} condition above, the QFI for estimating the normalized image moment $x_\mu$ using an $N$-telescope array satisfies
\begin{align}
\mbb{H}_{\mu\mu}(\rho_s)=
\begin{cases}
\Theta\bigl(\epsilon\Delta^{2\lceil \mu/2\rceil}\bigr), & \mu\le 2N-2,\\
O\bigl(\epsilon\Delta^{2\mu-2N+2}\bigr), & \mu> 2N-2.
\end{cases}
\label{eq:qfi-scaling-main-result}
\end{align}
The factor $\epsilon$ comes from the weight of the one-photon block in Eq.~\eqref{eq:rho}, i.e., $\mbb{H}_{\mu\mu}(\rho_s)=\epsilon\mbb{H}_{\mu\mu}(\rho_s^{(1)})+O(\epsilon^2)$. 
\end{theorem}

Thus, an $N$-telescope array reproduces the single-aperture
moment-estimation scaling up to the cutoff $\mu=2N-2$~\cite{Tsang2019} even without local mode sorting.


We now compare the full-array benchmark with pairwise strategies of the kind underlying conventional pairwise interferometry in the VCZ paradigm. By a \emph{pairwise measurement scheme}, we mean a receiver that allows an arbitrary quantum measurement on any two-telescope subspace $\mc H^{(1)}_{ij}=\mathrm{Span}\{\ket{1_i},\ket{1_j}\}$. Let $P_{ij}:=\op{1_i}{1_i}+\op{1_j}{1_j}$, and define the subnormalized pair-restricted state
$\mc P_{ij}[\rho_s]:=P_{ij}\rho_sP_{ij}$. We therefore define the normalized pairwise benchmark
\begin{align}
\mbb H^{\mathrm{(pair)}}_{\mu\mu}(\rho_s):=\max_{i<j} \mbb H_{\mu\mu}\left(\frac{\mc{P}_{ij}[\rho_s]}{\tr(\mc{P}_{ij}[\rho_s])}\right){\tr(\mc{P}_{ij}[\rho_s])},
\end{align}
where $\tr(\mc P_{ij}[\rho_s])=2\epsilon/N+O(\epsilon^2)$
is independent of the image moments to leading order.

This benchmark upper-bounds any randomized or adaptive receiver whose coherent measurement on each temporal mode has support on at most one telescope pair within the present single-mode-per-telescope model. However, it should not be interpreted as a bound on schemes that first extract additional intra-telescope spatial modes, since those degrees of freedom can also help estimate image moments and have been deliberately removed from the model.

Applying Thm.~\ref{Thm:qfi-scaling} with $N=2$ to each pair-restricted state gives the following bound.

\begin{theorem}
\label{thm:pairwise-qfi-bound}
Under the \textit{nondegeneracy} condition introduced above, the pairwise benchmark for estimating the normalized image moment $x_\mu$
obeys
\begin{align}
\mbb{H}^{\mathrm{(pair)}}_{\mu\mu}(\rho_s)
=
\begin{cases}
\Theta\bigl(\epsilon\Delta^{2\lceil \mu/2\rceil}\bigr), & \mu\le 2,\\
O\bigl(\epsilon\Delta^{2\mu-2}\bigr), & \mu> 2.
\end{cases}
\label{eq:pairwise-qfi-high}
\end{align}
Moreover, the Supplemental Material~\cite{SM} proves the sharper scaling for $\mu>2$, which gives
\begin{align}
\mbb{H}^{\mathrm{(pair)}}_{\mu\mu}(\rho_s)
=
\begin{cases}
\Theta\bigl(\epsilon\Delta^{2\mu}\bigr), & \mu \ \text{odd},\\
\Theta\bigl(\epsilon\Delta^{2\mu-2}\bigr), & \mu ~\text{even}.
\end{cases}
\label{eq:pairwise-qfi-parity}
\end{align}
\end{theorem}
\begin{remark}
This bound is consistent with earlier superresolution results for simple sources such as balanced incoherent two-point sources, whose local estimation problem is governed by low-order moments and can therefore be addressed by pairwise interference~\cite{lupo2020,wang2021,Sajjad2024, Padilla2026}. For $N\ge 3$, however, pairwise measurements are asymptotically
suboptimal for all higher moments
$3\le \mu\le 2N-2$ after isolating the inter-telescope degree of freedom. 
\end{remark}
\yk
For $N=3$, the full-array/pairwise separation already appears in third- and fourth-moment estimation.
Appendix~\ref{app:toy-example} gives an explicit toy source family whose unknown parameter $\eta$ first enters the fourth-order moment. A projection onto the array-SPADE mode
$(|1_1\rangle-2|1_2\rangle+|1_3\rangle)/\sqrt{6}$ (see Eq.~\eqref{eq:e2-three-telescope}) gives
$F_{\eta\eta}=\Theta(\epsilon\Delta^4)$, while Theorem 2 bounds every
pairwise strategy by $O(\epsilon\Delta^6)$. This gives an explicit instance in which quantum-limited moment estimation requires a measurement effect coherently supported on all three telescope modes. \blk

\text{FI scaling--} For a measurement $\mc M=\{M_r\}_r$ with outcome probabilities $p_r=\tr(\rho^{(1)}_s M_r)$, the Fisher information (FI) is $\mbb F_{\mu\mu}=\sum_r\frac{1}{p_r}\bigl(\partial_{x_\mu} p_r\bigr)^2$. 
Since $\mbb F_{\mu\mu}\le \mbb H_{\mu\mu}$ for every measurement, the QFI provides the fundamental benchmark for the attainable estimation scaling of $x_\mu$, and we now show that the QFI bound in Thm.~\ref{Thm:qfi-scaling} is tight in $\Delta$-scaling. 

\begin{proposition}
\label{prop:FI-saturates-QFI-scaling}
For every image moment $x_\mu$ with $\mu\le 2N-2$, there exists a measurement $\mc M_\mu$ such that
\begin{align}
\mbb F_{\mu\mu}^{(\mc M_\mu)}(\rho_s)
=
\Theta\bigl(\epsilon\Delta^{2\lceil \mu/2\rceil}\bigr),
\label{eq:FI-scaling}
\end{align}
Concretely, for even moments $\mu=2n$ with $n\le N-1$, the binary measurement $\mc M_{2n}=\{\op{e_n}{e_n}, \mbb 1-\op{e_n}{e_n}\}$ is sufficient. For odd moments $\mu=2n+1$ with $ n\le N-2$, one may use the three-outcome measurement $\mc M_{2n+1}=\left\{\op{\phi_{n,\pm}}{\phi_{n,\pm}},\mbb 1-\sum_{\pm}\op{\phi_{n,\pm}}{\phi_{n,\pm}}
\right\}$ built from $\ket{\phi_{n,\pm}}=\frac{\ket{e_n}\pm e^{i\chi_n}\ket{e_{n+1}}}{\sqrt{2}}$ with a suitable choice of phase $\chi_n$.
\end{proposition}

The proof is given in Appendix~\ref{app:proofs}, where we also show that all optimal scalings below the cutoff can be accessed using at most two projective measurement settings. In this sense, array-SPADE differs from architectures that first perform SPADE locally at each telescope and then interfere the sorted modes pairwise~\cite{Padilla2026,Padilla2026a}. Here, the array-SPADE measurement is itself distributed over the synthesized aperture: the inter-telescope interference implements the desired mode projection.

\textit{Quantum-network implementation.--} The preceding results identify the required measurement resource: rank-one one-photon effects with coherent support on three or more telescope modes. If the optical modes can be physically brought together through a low-loss channel, such projectors can be implemented directly. We now give two proof-of-principle distributed embeddings, not intended to optimize resource cost, showing that the same measurement primitives can be induced probabilistically in ancilla- and memory-assisted architectures, generalizing the original GJC and memory-assisted schemes~\cite{Gottesman2012,Khabiboulline2019PRL}, and show that the required full-array projections are compatible, at least in principle, with experimentally emerging quantum-network architectures~\cite{Stas2026, Wang2026}. 

To implement measurement effects with coherent support on multiple telescope modes, a natural first attempt is to use a one-photon $W$-state ancilla. However, a $W$-state ancilla does not overcome the pairwise restriction: after local postselection, the click pattern still identifies the two sites occupied by the source and ancilla photons, so the induced effect remains effectively pairwise~\cite{Gottesman2012}. Here, we instead use an $(N-1)$-photon $M$ state to give a simple postselected construction.

Consider the $(N-1)$-photon $M$ state
\begin{subequations}
\begin{align}
&\ket{\phi}^{M}_{a}=\frac{1}{\sqrt{N}}\sum_j\ket{1_1\cdots 0_j\cdots 1_N}_a,\\
&\ket{u_j}_{s_j a_j}=t_j\ket{0}_{s_j}\ket{1}_{a_j}+
\sqrt{1-|t_j|^2}\ket{1}_{s_j}\ket{0}_{a_j},
\end{align}
\end{subequations}
where $s_j$ and $a_j$ are the signal and ancilla modes on telescope $j$. Postselecting on successful local projections onto $\ket{u_j}$ at every telescope induces a rank-one effect $\op{w}{w}_s$ on the one-photon source subspace $\mc H^{(1)}_s$ with $\bra{w}_s=\frac{1}{\sqrt N}\sum_j
\left(
\sqrt{1-|t_j|^2}\prod_{k\neq j}t^*_k
\right)\bra{1_j}_s$. Hence, by varying the complex amplitude $\{t_j\}$, one can probabilistically realize an arbitrary rank-one projector on $\mc H^{(1)}_s$ with probability $p_{\mathrm{succ}}=\frac{1}{N}\sum_j (1-|t_j|^2)\prod_{k\ne j}|t_k|^2$.

\begin{figure}
    \centering
    \includegraphics[width=0.9\linewidth]{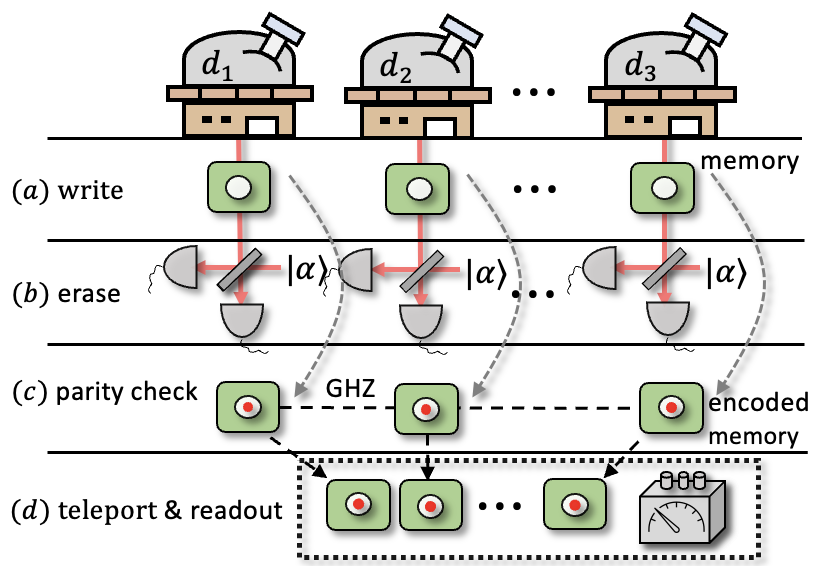}
    \caption{Memory-assisted implementation of a genuine three-telescope measurement: (a) signal photon collection with local memories; (b) heralded photon erasure to complete photon state storage using local oscillators; (c) removal of the vacuum component through parity check with GHZ state; (d) teleportation and central node readout.  The scheme is a proof-of-principle representation of the array-SPADE measurement primitives identified in the main text.}
    \label{fig:assisted}
\end{figure}

The same full-array measurement primitives can also be embedded into a
memory-assisted scheme~\cite{Stas2026, Wang2026}.  

Fig.~\ref{fig:assisted} sketches a memory-assisted route that generalizes the recent memory-assisted pairwise quantum-telescope architecture~\cite{Stas2026, Wang2026}.
The incoming one-photon array state is first coherently written into distributed memories with the herald record erased using local oscillators and the vacuum component removed by a nonlocal parity check. The resulting memory qudit is then measured by a central or distributed readout. A detailed proof-of-principle construction is provided in the Supplemental Material~\cite{SM}. Its purpose is to show that the genuinely multi-telescope measurement resource required by the QFI analysis is compatible, in principle, with realistic memory-assisted
quantum-telescope architectures.

\textit{Conclusion.--} We have derived the Quantum Fisher Information (QFI) scaling for estimating the normalized image moments of an extended source observed by an $N$-telescope array in the weak-light, subdiffraction regime. In a single-mode-per-telescope model, full-array measurements reproduce the single-aperture moment-estimation scaling up to the finite-array cutoff $2N-2$. By contrast, arbitrary measurements restricted to telescope pairs attain this scaling only up to second order. Thus, beyond the first- and second-order moments, quantum-limited subdiffraction moment estimation requires genuinely multi-telescope interference.

We constructed array-SPADE measurements that attain the full-array scaling. These measurements are synthesized-aperture analogues of SPADE: they are not local spatial-mode sorters performed separately at each telescope, but coherent one-photon projections across the telescope array. We further showed that the corresponding measurement primitives can, in principle, be embedded in ancilla- and memory-assisted quantum-telescope architectures. Thus, in quantum-limited long-baseline subdiffraction moment estimation, the essential resource is not merely access to long baselines or pairwise visibilities, but the ability to perform coherent one-photon projections across the full telescope array.

The scalings established here are local, diagonal QFI/FI scalings for individual normalized image moments. They are sufficient to show an information-theoretic scaling gap between pairwise-restricted receivers and genuinely multi-telescope interference. A full multiparameter analysis of the simultaneous estimation of several source moments, including effects beyond the leading-order scaling established here, remains an important direction for future work.

\textit{Acknowledgments.--}The research has been conducted at the Institute for Quantum Computing, University of Waterloo, which is supported by Innovation, Science, and Economic Development Canada. This work is co-funded by the Natural Sciences and Engineering Research Council of Canada (NSERC) and the European Union (EU) under Grant No. 101070168 (HyperSpace), and the Canada Excellence Research Chair (CERC) program. WW acknowledges support from the NSERC CREATE in Quantum Computing Program, grant number 543245. YW acknowledges funding from the Canada First Research Excellence Fund, NSERC-NSF Alliance Grant No. ALLRP-586858-2023, and NSERC-UKRI Alliance Grant No. ALLRP-597823-24.\blk

\bibliography{ref}

@article{Gottesman2012,
  title = {Longer-Baseline Telescopes Using Quantum Repeaters},
  author = {Gottesman, Daniel and Jennewein, Thomas and Croke, Sarah},
  journal = {Physical Review Letters},
  volume = {109},
  number = {7},
  pages = {070503},
  year = {2012},
  doi = {10.1103/PhysRevLett.109.070503}
}

@article{Khabiboulline2019PRL,
  title = {Optical Interferometry with Quantum Networks},
  author = {Khabiboulline, Emil T. and Borregaard, Johannes and De Greve, Kristiaan and Lukin, Mikhail D.},
  journal = {Physical Review Letters},
  volume = {123},
  number = {7},
  pages = {070504},
  year = {2019},
  doi = {10.1103/PhysRevLett.123.070504}
}

@article{Khabiboulline2019PRA,
  title = {Quantum-Assisted Telescope Arrays},
  author = {Khabiboulline, Emil T. and Borregaard, Johannes and De Greve, Kristiaan and Lukin, Mikhail D.},
  journal = {Physical Review A},
  volume = {100},
  number = {2},
  pages = {022316},
  year = {2019},
  doi = {10.1103/PhysRevA.100.022316}
}

@article{Tsang2016,
  title = {Quantum Theory of Superresolution for Two Incoherent Optical Point Sources},
  author = {Tsang, Mankei and Nair, Ranjith and Lu, Xiao-Ming},
  journal = {Physical Review X},
  volume = {6},
  number = {3},
  pages = {031033},
  year = {2016},
  doi = {10.1103/PhysRevX.6.031033}
}

@article{Nair2016,
  title = {Far-Field Superresolution of Thermal Electromagnetic Sources at the Quantum Limit},
  author = {Nair, Ranjith and Tsang, Mankei},
  journal = {Physical Review Letters},
  volume = {117},
  number = {19},
  pages = {190801},
  year = {2016},
  doi = {10.1103/PhysRevLett.117.190801}
}

@article{Tsang2017,
  title = {Subdiffraction Incoherent Optical Imaging via Spatial-Mode Demultiplexing},
  author = {Tsang, Mankei},
  journal = {New Journal of Physics},
  volume = {19},
  number = {2},
  pages = {023054},
  year = {2017},
  doi = {10.1088/1367-2630/aa60ee}
}

@article{Tsang2018,
  title = {Subdiffraction Incoherent Optical Imaging via Spatial-Mode Demultiplexing: Semiclassical Treatment},
  author = {Tsang, Mankei},
  journal = {Physical Review A},
  volume = {97},
  number = {2},
  pages = {023830},
  year = {2018},
  doi = {10.1103/PhysRevA.97.023830}
}

@article{Cittert1934,
  author = {van Cittert, P. H.},
  title = {Die Wahrscheinliche Schwingungsverteilung in Einer von Einer Lichtquelle Direkt Oder Mittels Einer Linse Beleuchteten Ebene},
  journal = {Physica},
  volume = {1},
  pages = {201--210},
  year = {1934},
  doi = {10.1016/S0031-8914(34)90027-6}
}

@book{Goodman2017,
  author = {Goodman, Joseph W.},
  title = {Introduction to Fourier Optics},
  edition = {4},
  publisher = {W. H. Freeman},
  year = {2017},
  isbn = {9781319119164}
}

@article{Monnier2003,
  author = {Monnier, John D.},
  title = {Optical Interferometry in Astronomy},
  journal = {Reports on Progress in Physics},
  volume = {66},
  number = {5},
  pages = {789--857},
  year = {2003},
  doi = {10.1088/0034-4885/66/5/203}
}

@article{Wang2025,
  title = {Limitations of Gaussian Measurements in Quantum Imaging},
  author = {Wang, Yunkai and Zhou, Sisi},
  journal = {Phys. Rev. Lett.},
  volume = {135},
  issue = {12},
  pages = {120201},
  numpages = {7},
  year = {2025},
  month = {Sep},
  publisher = {American Physical Society},
  doi = {10.1103/hnhp-jhr2},
  url = {https://link.aps.org/doi/10.1103/hnhp-jhr2}
}

@article{Wang2025b,
  title = {Temporally Localized Quantum Operations on Continuous-Wave Thermal Light},
  author = {Wang, Yunkai and Zhang, Yujie and Lorenz, Virginia O.},
  journal = {Phys. Rev. Lett.},
  volume = {135},
  issue = {11},
  pages = {113602},
  numpages = {6},
  year = {2025},
  month = {Sep},
  publisher = {American Physical Society},
  doi = {10.1103/y4v8-1wgm},
  url = {https://link.aps.org/doi/10.1103/y4v8-1wgm}
}

@Article{Wang2025d,
author={Wang, Yunkai
and Oh, Changhun
and Liu, Junyu
and Jiang, Liang
and Zhou, Sisi},
title={Advancing quantum imaging through learning theory},
journal={Nature Communications},
year={2025},
month={Dec},
day={27},
volume={17},
number={1},
pages={1124},
issn={2041-1723},
doi={10.1038/s41467-025-67884-1},
url={https://doi.org/10.1038/s41467-025-67884-1}
}

@Article{Zanforlin2022,
author={Zanforlin, Ugo
and Lupo, Cosmo
and Connolly, Peter W. R.
and Kok, Pieter
and Buller, Gerald S.
and Huang, Zixin},
title={Optical quantum super-resolution imaging and hypothesis testing},
journal={Nature Communications},
year={2022},
month={Sep},
day={13},
volume={13},
number={1},
pages={5373},
issn={2041-1723},
doi={10.1038/s41467-022-32977-8},
url={https://doi.org/10.1038/s41467-022-32977-8}
}

@article{Tsang2019review,
author = {Mankei Tsang},
title = {Resolving starlight: a quantum perspective},
journal = {Contemporary Physics},
volume = {60},
number = {4},
pages = {279--298},
year = {2019},
publisher = {Taylor \& Francis},
doi = {10.1080/00107514.2020.1736375},
URL = {https://doi.org/10.1080/00107514.2020.1736375
},
eprint = { https://doi.org/10.1080/00107514.2020.1736375}
}

@article{Matthew2023,
  title = {Interferometric Imaging Using Shared Quantum Entanglement},
  author = {Brown, Matthew R. and Allgaier, Markus and Thiel, Val\'erian and Monnier, John D. and Raymer, Michael G. and Smith, Brian J.},
  journal = {Phys. Rev. Lett.},
  volume = {131},
  issue = {21},
  pages = {210801},
  numpages = {6},
  year = {2023},
  month = {Nov},
  publisher = {American Physical Society},
  doi = {10.1103/PhysRevLett.131.210801},
  url = {https://link.aps.org/doi/10.1103/PhysRevLett.131.210801}
}

@article{Wang2021,
  title = {Superresolution in interferometric imaging of strong thermal sources},
  author = {Wang, Yunkai and Zhang, Yujie and Lorenz, Virginia O.},
  journal = {Phys. Rev. A},
  volume = {104},
  issue = {2},
  pages = {022613},
  numpages = {11},
  year = {2021},
  month = {Aug},
  publisher = {American Physical Society},
  doi = {10.1103/PhysRevA.104.022613},
  url = {https://link.aps.org/doi/10.1103/PhysRevA.104.022613}
}

@article{Ang2017,
  title = {Quantum limit for two-dimensional resolution of two incoherent optical point sources},
  author = {Ang, Shan Zheng and Nair, Ranjith and Tsang, Mankei},
  journal = {Phys. Rev. A},
  volume = {95},
  issue = {6},
  pages = {063847},
  numpages = {13},
  year = {2017},
  month = {Jun},
  publisher = {American Physical Society},
  doi = {10.1103/PhysRevA.95.063847},
  url = {https://link.aps.org/doi/10.1103/PhysRevA.95.063847}
}

@article{Wang2026,
  title = {Memory-Assisted Nonlocal Interferometer toward Long-Baseline Telescopes},
  author = {Wang, Bin and Luo, Xi-Yu and Gao, Bo-Feng and Liu, Jian-Long and Wang, Chao-Yang and Yan, Zi and Ke, Qiao-Mu and Teng, Da and Zheng, Ming-Yang and Cao, Yuan and Li, Jun and Peng, Cheng-Zhi and Zhang, Qiang and Bao, Xiao-Hui and Pan, Jian-Wei},
  journal = {Phys. Rev. Lett.},
  volume = {136},
  issue = {24},
  pages = {240801},
  numpages = {7},
  year = {2026},
  month = {Jun},
  publisher = {American Physical Society},
  doi = {10.1103/qpzn-h7p9},
  url = {https://link.aps.org/doi/10.1103/qpzn-h7p9}
}

@article{Sorelli2021a,
  title = {Optimal Observables and Estimators for Practical Superresolution Imaging},
  author = {Sorelli, Giacomo and Gessner, Manuel and Walschaers, Mattia and Treps, Nicolas},
  journal = {Phys. Rev. Lett.},
  volume = {127},
  issue = {12},
  pages = {123604},
  numpages = {6},
  year = {2021},
  month = {Sep},
  publisher = {American Physical Society},
  doi = {10.1103/PhysRevLett.127.123604},
  url = {https://link.aps.org/doi/10.1103/PhysRevLett.127.123604}
}

@article{Sorelli2021b,
  title = {Moment-based superresolution: Formalism and applications},
  author = {Sorelli, Giacomo and Gessner, Manuel and Walschaers, Mattia and Treps, Nicolas},
  journal = {Phys. Rev. A},
  volume = {104},
  issue = {3},
  pages = {033515},
  numpages = {18},
  year = {2021},
  month = {Sep},
  publisher = {American Physical Society},
  doi = {10.1103/PhysRevA.104.033515},
  url = {https://link.aps.org/doi/10.1103/PhysRevA.104.033515}
}

@article{Narayan1986MaxEntAstronomy,
  author  = {Narayan, Ramesh and Nityananda, Rajaram},
  title   = {Maximum Entropy Image Restoration in Astronomy},
  journal = {Annual Review of Astronomy and Astrophysics},
  volume  = {24},
  number  = {1},
  pages   = {127--170},
  year    = {1986},
  doi     = {10.1146/annurev.aa.24.090186.001015}
}

@article{Thiebaut2013Interferometry,
  author  = {Thi{\'e}baut, {\'E}ric},
  title   = {Principles of Image Reconstruction in Interferometry},
  journal = {EAS Publications Series},
  volume  = {59},
  pages   = {157--187},
  year    = {2013},
  doi     = {10.1051/eas/1359009}
}

@Article{Olmi2026,
author={Olmi, Luca
and Migoni, Carlo
and Murgia, Matteo
and Nesti, Renzo
and Poppi, Sergio},
title={First demonstration of super-resolution with a single-aperture radio telescope},
journal={Experimental Astronomy},
year={2026},
month={Mar},
day={11},
volume={61},
number={2},
pages={5},
issn={1572-9508},
doi={10.1007/s10686-025-10041-9},
url={https://doi.org/10.1007/s10686-025-10041-9}
}

@article{Dutton2019,
  title = {Attaining the quantum limit of superresolution in imaging an object's length via predetection spatial-mode sorting},
  author = {Dutton, Zachary and Kerviche, Ronan and Ashok, Amit and Guha, Saikat},
  journal = {Phys. Rev. A},
  volume = {99},
  issue = {3},
  pages = {033847},
  numpages = {8},
  year = {2019},
  month = {Mar},
  publisher = {American Physical Society},
  doi = {10.1103/PhysRevA.99.033847},
  url = {https://link.aps.org/doi/10.1103/PhysRevA.99.033847}
}

@article{Tsang2018a,
  title = {Conservative classical and quantum resolution limits for incoherent imaging},
  author = {Tsang, Mankei},
  journal = {Journal of Modern Optics},
  volume = {65},
  number = {11},
  pages = {1385--1391},
  year = {2018},
  doi = {10.1080/09500340.2017.1377306}
}

@article{lupo2020,
  title = {Quantum Limits to Incoherent Imaging are Achieved by Linear Interferometry},
  author = {Lupo, Cosmo and Huang, Zixin and Kok, Pieter},
  journal = {Phys. Rev. Lett.},
  volume = {124},
  issue = {8},
  pages = {080503},
  numpages = {6},
  year = {2020},
  month = {Feb},
  publisher = {American Physical Society},
  doi = {10.1103/PhysRevLett.124.080503},
  url = {https://link.aps.org/doi/10.1103/PhysRevLett.124.080503}
}

@article{Padilla2026,
  title = {Superresolution Imaging with Entanglement-Enhanced Telescopy},
  author = {Padilla, Isack and Sajjad, Aqil and Saif, Babak N. and Guha, Saikat},
  journal = {Phys. Rev. Lett.},
  volume = {136},
  issue = {1},
  pages = {010803},
  numpages = {7},
  year = {2026},
  month = {Jan},
  publisher = {American Physical Society},
  doi = {10.1103/354q-ch63},
  url = {https://link.aps.org/doi/10.1103/354q-ch63}
}

@article{Padilla2026a,
  title = {Quantum resolution limit of long-baseline imaging using distributed entanglement},
  author = {Padilla, Isack and Sajjad, Aqil and Saif, Babak N. and Guha, Saikat},
  journal = {Phys. Rev. A},
  volume = {113},
  issue = {1},
  pages = {012608},
  numpages = {21},
  year = {2026},
  month = {Jan},
  publisher = {American Physical Society},
  doi = {10.1103/4npc-gpvh},
  url = {https://link.aps.org/doi/10.1103/4npc-gpvh}
}

@article{Sajjad2024,
  title = {Quantum limits of parameter estimation in long-baseline imaging},
  author = {Sajjad, Aqil and Grace, Michael R. and Guha, Saikat},
  journal = {Phys. Rev. Res.},
  volume = {6},
  issue = {1},
  pages = {013212},
  numpages = {25},
  year = {2024},
  month = {Feb},
  publisher = {American Physical Society},
  doi = {10.1103/PhysRevResearch.6.013212},
  url = {https://link.aps.org/doi/10.1103/PhysRevResearch.6.013212}
}

@article{Huang2026,
author = {Zixin Huang and Oleg Titov and Mikołaj K. Schmidt and Benjamin Pope and Gavin K. Brennen and Daniel K. L. Oi and Pieter Kok},
title = {Quantum-enabled optical large-baseline interferometry: applications, protocols and feasibility},
journal = {Advances in Physics: X},
volume = {11},
number = {1},
pages = {2597311},
year = {2026},
publisher = {Taylor \& Francis},
doi = {10.1080/23746149.2025.2597311},
URL = {https://doi.org/10.1080/23746149.2025.2597311},
eprint = {https://doi.org/10.1080/23746149.2025.2597311}
}

@inproceedings{Diaz21,
author = {David Diaz and Yujie Zhang and Virginia O. Lorenz and Paul G. Kwiat},
booktitle = {Frontiers in Optics $+$ Laser Science 2021},
journal = {Frontiers in Optics $+$ Laser Science 2021},
pages = {FTh6D.7},
publisher = {Optica Publishing Group},
title = {Emulating Quantum-enhanced Long-Baseline Interferometric Telescopy},
year = {2021},
url = {https://opg.optica.org/abstract.cfm?URI=FiO-2021-FTh6D.7},
doi = {10.1364/FIO.2021.FTh6D.7},
}

@article{Zhang2025,
  title = {Criteria for optimal entanglement-assisted long baseline telescopy},
  author = {Zhang, Yujie and Jennewein, Thomas},
  journal = {Phys. Rev. Res.},
  volume = {7},
  issue = {4},
  pages = {043278},
  numpages = {20},
  year = {2025},
  month = {Dec},
  publisher = {American Physical Society},
  doi = {10.1103/bf51-tj3j},
  url = {https://link.aps.org/doi/10.1103/bf51-tj3j}
}

@article{ZhouJiang2019,
  title = {Modern Description of Rayleigh's Criterion},
  author = {Zhou, Sisi and Jiang, Liang},
  journal = {Physical Review A},
  volume = {99},
  number = {1},
  pages = {013808},
  year = {2019},
  doi = {10.1103/PhysRevA.99.013808}
}

@article{Tsang2019Semiparametric,
  title = {Semiparametric Estimation for Incoherent Optical Imaging},
  author = {Tsang, Mankei},
  journal = {Physical Review Research},
  volume = {1},
  number = {3},
  pages = {033006},
  year = {2019},
  doi = {10.1103/PhysRevResearch.1.033006}
}

@article{Tsang2019 ,
  title = {Quantum limit to subdiffraction incoherent optical imaging},
  author = {Tsang, Mankei},
  journal = {Phys. Rev. A},
  volume = {99},
  issue = {1},
  pages = {012305},
  numpages = {12},
  year = {2019},
  month = {Jan},
  publisher = {American Physical Society},
  doi = {10.1103/PhysRevA.99.012305},
  url = {https://link.aps.org/doi/10.1103/PhysRevA.99.012305}
}

@Article{Stas2026,
author={Stas, P.-J.
and Wei, Y.-C.
and Sirotin, M.
and Huan, Y. Q.
and Yazlar, U.
and Abdo Arias, F.
and Knyazev, E.
and Baranes, G.
and Machielse, B.
and Grandi, S.
and Riedel, D.
and Borregaard, J.
and Park, H.
and Lon{\v{c}}ar, M.
and Suleymanzade, A.
and Lukin, M. D.},
title={Entanglement-assisted non-local optical interferometry in a quantum network},
journal={Nature},
year={2026},
month={Mar},
day={01},
volume={651},
number={8105},
pages={326-332},
issn={1476-4687},
doi={10.1038/s41586-026-10171-w},
url={https://doi.org/10.1038/s41586-026-10171-w}
}

@article{Czupryniak2023,
  title = {Optimal qubit circuits for quantum-enhanced telescopes},
  author = {Czupryniak, Robert and Steinmetz, John and Kwiat, Paul G. and Jordan, Andrew N.},
  journal = {Phys. Rev. A},
  volume = {108},
  issue = {5},
  pages = {052408},
  numpages = {13},
  year = {2023},
  month = {Nov},
  publisher = {American Physical Society},
  doi = {10.1103/PhysRevA.108.052408},
  url = {https://link.aps.org/doi/10.1103/PhysRevA.108.052408}
}

@article{wang2023astronomical,
  title = {Astronomical interferometry using continuous variable quantum teleportation},
  author = {Wang, Yunkai and Zhang, Yujie and Lorenz, Virginia O.},
  journal = {Phys. Rev. Res.},
  volume = {7},
  issue = {2},
  pages = {023154},
  numpages = {15},
  year = {2025},
  month = {May},
  publisher = {American Physical Society},
  doi = {10.1103/PhysRevResearch.7.023154},
  url = {https://link.aps.org/doi/10.1103/PhysRevResearch.7.023154}
}

@article{Wang21,
author = {Ben Wang and Liang Xu and Jun-chi Li and Lijian Zhang},
journal = {Photon. Res.},
number = {8},
pages = {1522--1530},
publisher = {Optica Publishing Group},
title = {Quantum-limited localization and resolution in three dimensions},
volume = {9},
month = {Aug},
year = {2021},
url = {https://opg.optica.org/prj/abstract.cfm?URI=prj-9-8-1522},
doi = {10.1364/PRJ.417613},
}

@article{Ou2021,
  title = {Quantum Limits of Superresolution in a Noisy Environment},
  author = {Oh, Changhun and Zhou, Sisi and Wong, Yat and Jiang, Liang},
  journal = {Phys. Rev. Lett.},
  volume = {126},
  issue = {12},
  pages = {120502},
  numpages = {7},
  year = {2021},
  month = {Mar},
  publisher = {American Physical Society},
  doi = {10.1103/PhysRevLett.126.120502},
  url = {https://link.aps.org/doi/10.1103/PhysRevLett.126.120502}
}

@article{Paur16,
author = {Martin Pa\'{u}r and Bohumil Stoklasa and Zdenek Hradil and Luis L. S\'{a}nchez-Soto and Jaroslav Rehacek},
journal = {Optica},
number = {10},
pages = {1144--1147},
publisher = {Optica Publishing Group},
title = {Achieving the ultimate optical resolution},
volume = {3},
month = {Oct},
year = {2016},
url = {https://opg.optica.org/optica/abstract.cfm?URI=optica-3-10-1144},
doi = {10.1364/OPTICA.3.001144},
}

@article{Tham2017,
  title = {Beating Rayleigh's Curse by Imaging Using Phase Information},
  author = {Tham, Weng-Kian and Ferretti, Hugo and Steinberg, Aephraim M.},
  journal = {Phys. Rev. Lett.},
  volume = {118},
  issue = {7},
  pages = {070801},
  numpages = {6},
  year = {2017},
  month = {Feb},
  publisher = {American Physical Society},
  doi = {10.1103/PhysRevLett.118.070801},
  url = {https://link.aps.org/doi/10.1103/PhysRevLett.118.070801}
}

@article{Wang2023,
  title = {Fundamental limit of bandwidth-extrapolation-based superresolution},
  author = {Wang, Yunkai and Lorenz, Virginia O.},
  journal = {Phys. Rev. A},
  volume = {108},
  issue = {1},
  pages = {012602},
  numpages = {12},
  year = {2023},
  month = {Jul},
  publisher = {American Physical Society},
  doi = {10.1103/PhysRevA.108.012602},
  url = {https://link.aps.org/doi/10.1103/PhysRevA.108.012602}
}

@article{Hu2025,
  title = {Superresolution of unequal-brightness thermal sources for stellar interferometry},
  author = {Hu, Chenyu and Wang, Ben and Zhang, Jiandong and Wang, Kunxu and Liu, Huigen and Zhou, Jilin and Zhang, Lijian},
  journal = {Phys. Rev. A},
  volume = {112},
  issue = {3},
  pages = {032609},
  numpages = {11},
  year = {2025},
  month = {Sep},
  publisher = {American Physical Society},
  doi = {10.1103/z157-q6n3},
  url = {https://link.aps.org/doi/10.1103/z157-q6n3}
}

@article{Lupo2016,
  title = {Ultimate Precision Bound of Quantum and Subwavelength Imaging},
  author = {Lupo, Cosmo and Pirandola, Stefano},
  journal = {Phys. Rev. Lett.},
  volume = {117},
  issue = {19},
  pages = {190802},
  numpages = {5},
  year = {2016},
  month = {Nov},
  publisher = {American Physical Society},
  doi = {10.1103/PhysRevLett.117.190802},
  url = {https://link.aps.org/doi/10.1103/PhysRevLett.117.190802}
}

@article{Marchese2023,
  title = {Large Baseline Optical Imaging Assisted by Single Photons and Linear Quantum Optics},
  author = {Marchese, Marta Maria and Kok, Pieter},
  journal = {Phys. Rev. Lett.},
  volume = {130},
  issue = {16},
  pages = {160801},
  numpages = {6},
  year = {2023},
  month = {Apr},
  publisher = {American Physical Society},
  doi = {10.1103/PhysRevLett.130.160801},
  url = {https://link.aps.org/doi/10.1103/PhysRevLett.130.160801}
}

@article{tsang2011,
  title = {Quantum Nonlocality in Weak-Thermal-Light Interferometry},
  author = {Tsang, Mankei},
  journal = {Phys. Rev. Lett.},
  volume = {107},
  issue = {27},
  pages = {270402},
  numpages = {5},
  year = {2011},
  month = {Dec},
  publisher = {American Physical Society},
  doi = {10.1103/PhysRevLett.107.270402},
  url = {https://link.aps.org/doi/10.1103/PhysRevLett.107.270402}
}

@article{Zernike1938,
title = {The concept of degree of coherence and its application to optical problems},
journal = {Physica},
volume = {5},
number = {8},
pages = {785-795},
year = {1938},
issn = {0031-8914},
doi = {https://doi.org/10.1016/S0031-8914(38)80203-2},
url = {https://www.sciencedirect.com/science/article/pii/S0031891438802032},
author = {F. Zernike}
}

@article{huang2024,
  title = {Limited quantum advantage for stellar interferometry via continuous-variable teleportation},
  author = {Huang, Zixin and Baragiola, Ben Q. and Menicucci, Nicolas C. and Wilde, Mark M.},
  journal = {Phys. Rev. A},
  volume = {109},
  issue = {5},
  pages = {052434},
  numpages = {14},
  year = {2024},
  month = {May},
  publisher = {American Physical Society},
  doi = {10.1103/PhysRevA.109.052434},
  url = {https://link.aps.org/doi/10.1103/PhysRevA.109.052434}
}

@article{Huang2022,
  title = {Imaging Stars with Quantum Error Correction},
  author = {Huang, Zixin and Brennen, Gavin K. and Ouyang, Yingkai},
  journal = {Phys. Rev. Lett.},
  volume = {129},
  issue = {21},
  pages = {210502},
  numpages = {6},
  year = {2022},
  month = {Nov},
  publisher = {American Physical Society},
  doi = {10.1103/PhysRevLett.129.210502},
  url = {https://link.aps.org/doi/10.1103/PhysRevLett.129.210502}
}

@misc{SM,
  title = {Supplementary Material},
  author = {Yujie Zhang}
}

@article{Tsang2021,
  title = {Quantum limit to subdiffraction incoherent optical imaging. II. A parametric-submodel approach},
  author = {Tsang, Mankei},
  journal = {Phys. Rev. A},
  volume = {104},
  issue = {5},
  pages = {052411},
  numpages = {16},
  year = {2021},
  month = {Nov},
  publisher = {American Physical Society},
  doi = {10.1103/PhysRevA.104.052411},
  url = {https://link.aps.org/doi/10.1103/PhysRevA.104.052411}
}

@article{Tan2023,
  title = {Quantum limit to subdiffraction incoherent optical imaging. III. Numerical analysis},
  author = {Tan, Xiao-Jie and Tsang, Mankei},
  journal = {Phys. Rev. A},
  volume = {108},
  issue = {5},
  pages = {052416},
  numpages = {8},
  year = {2023},
  month = {Nov},
  publisher = {American Physical Society},
  doi = {10.1103/PhysRevA.108.052416},
  url = {https://link.aps.org/doi/10.1103/PhysRevA.108.052416}
}
\appendix
\section{Proofs of Thm.~\ref{Thm:qfi-scaling}
and Prop.~\ref{prop:FI-saturates-QFI-scaling}}
\label{app:proofs}
Let $\{\ket{e_a}\}_{a=0}^{N-1}$ be the array-SPADE basis defined in Eq.~\eqref{GSbasis}.  Since this basis is independent of $\Delta$, 
\begin{align}
\bra{e_a}H^m\ket{\psi}
=
\begin{cases}
0, & m<a,\\
\Theta(1), & m=a,\\
O(1), & m>a.
\end{cases}
\label{eq:app-H-overlap}
\end{align}
It then follows directly from Eq.~\eqref{eq:A-mu-prelim} that
\begin{align}
\bra{e_a}A_\mu\ket{e_b}
=
\begin{cases}
0, & \mu<a+b,\\
\Theta(\Delta^\mu), & \mu=a+b,\\
O(\Delta^\mu), & \mu>a+b .
\end{cases}
\label{eq:app-A-scaling}
\end{align}

Since $\rho_s^{(1)}=\sum_{\mu\ge0}x_\mu A_\mu$, its matrix elements, written in the array-SPADE basis, have the form
\begin{align}
\bra{e_a}\rho_s^{(1)}\ket{e_b}=\Delta^{a+b}
\left(f_a x_{a+b} f_b^*+O(\Delta)\right),
\label{eq:rhoea}
\end{align}
where $f_a=\frac{(-i)^a}{a!}\bra{e_a}H^a\ket{\psi}=\Theta(1)$.  Equivalently, 
\begin{align}
\rho_s^{(1)}=D X D,\quad
D=\mathrm{diag}(1,\Delta,\ldots,\Delta^{N-1})
\end{align}
in the $\{\ket{e_a}\}$ basis, with
\begin{align}
X=F\Gamma F^\dagger+O(\Delta).
\end{align}
Here $F=\mathrm{diag}(f_0,\ldots,f_{N-1})$ and
$\Gamma_{ab}=x_{a+b}$.  Since $f_a=\Theta(1)$ and $\Gamma$ is positive definite by assumption, $X$ is uniformly positive and bounded for sufficiently small $\Delta$.  Hence, there exist constants
$c,C>0$ such that
\begin{align}
c\|D\ket v\|^2 \le \bra v\rho_s^{(1)}\ket v\le C\|D\ket v\|^2 .
\label{eq:app-quadratic}
\end{align}
By the min-max theorem, the ordered eigenvalues of $\rho_s^{(1)}$ then satisfy
\begin{align}
\lambda_q=\Theta(\Delta^{2q}),
\qquad q=0,\ldots,N-1.
\label{eq:app-lambda}
\end{align}
Now expanding $\ket q=\sum_{a=0}^{N-1}\braket{e_a|q}\ket{e_a}$ and applying Eq.~\eqref{eq:app-quadratic} to $\ket{q}$, one has
\begin{align}
&\|D\ket q\|^2=\sum_a \Delta^{2a}|\braket{e_a|q}|^2=\Theta(\Delta^{2q})\notag \\
&\Rightarrow|\braket{e_a|q}|= O(\Delta^{q-a}), \qquad a<q.
\label{eq:app-left-overlap}
\end{align}
Similarly, since $X^{-1}$ is also uniformly positive and bounded, we have $(\rho_s^{(1)})^{-1}=D^{-1}X^{-1}D^{-1}$ and 
\begin{align}
c\|D^{-1}\ket v\|^2\le\bra v(\rho_s^{(1)})^{-1}\ket v\le C\|D^{-1}\ket v\|^2 .
\label{eq:app-D-quadratic2}
\end{align}
Applying the same argument to the inverse gives
\begin{align}
&\|D^{-1}\ket q\|^2=\sum_a\Delta^{-2a}|\braket{e_a|q}|^2=\Theta(\Delta^{-2q}) \notag \\
&\Rightarrow |\braket{e_a|q}|=O(\Delta^{a-q}), \qquad  a>q. \label{eq:app-right-overlap}
\end{align}
Combining Eqs.~\eqref{eq:app-left-overlap},
\eqref{eq:app-right-overlap}, and the normalization condition $\sum_a |\braket{e_a|q}|^2=1$ then yields
\begin{align}
|\braket{e_a|q}|&=O(\Delta^{|a-q|}),\qquad a\ne q,\notag\\
|\braket{e_q|q}|&=\Theta(1).
\label{eq:app-overlap}
\end{align}

\paragraph*{Proof of Thm.~\ref{Thm:qfi-scaling} and Prop.~\ref{prop:FI-saturates-QFI-scaling}.}

We first prove an upper bound on the one-photon QFI,  and then obtain the matching lower bound from an explicit array-SPADE measurement. This establishes both Thm.~\ref{Thm:qfi-scaling} and Prop.~\ref{prop:FI-saturates-QFI-scaling}.

The one-photon QFI of the image moment $x_{\mu}$ is
\begin{align}
\mbb H_{\mu\mu}(\rho_s^{(1)})=2\sum_{p,q}\frac{|\bra p A_\mu\ket q|^2}{\lambda_p+\lambda_q}. \label{eq:app-QFI}
\end{align}
Equations~\eqref{eq:app-A-scaling} and \eqref{eq:app-overlap} imply
\begin{align}
&\left|\bra{p}A_\mu\ket{q}\right|
\le
\sum_{a+b\le \mu}
\left|\braket{p|e_a}\right|
\left|\bra{e_a}A_\mu\ket{e_b}\right|
\left|\braket{e_b|q}\right| \\
&=
\sum_{a+b\le \mu}
O\left(\Delta^{\mu+|a-p|+|b-q|}\right) =
O\left(\Delta^{\mu+\max\{0,p+q-\mu\}}\right),\notag
\end{align}
where in the last step, we use $\min_{a+b\le\mu} (|a-p|+|b-q|)=\max\{0,p+q-\mu\}$. Moreover, Eq.~\eqref{eq:app-lambda} gives
\begin{align}
\lambda_p+\lambda_q
=
\Theta(\Delta^{2\min\{p,q\}}).
\end{align}
Thus each summand in the QFI in Eq.~\eqref{eq:app-QFI} is bounded by $O\left(\Delta^{2\mu+2\max\{0,p+q-\mu\}-2\min\{p,q\}}\right)$, and the smallest exponent is obtained at $\min\{\lfloor\mu/2\rfloor,N-1\}$, so
\begin{align}
\mbb H_{\mu\mu}(\rho_s^{(1)})=O(\Delta^{2\mu-2\min\{\lfloor\mu/2\rfloor,N-1\}}).
\label{eq:app-qfi-upper}
\end{align}

It remains to prove the matching lower bound for $\mu\le 2N-2$.   We do so by constructing an explicit measurement whose Fisher information has the desired scaling. Since the classical Fisher information of any measurement is upper-bounded by the QFI, this also lower-bounds QFI.

For $\mu=2n$, one can use the binary projective measurement
$\{\op{e_n}{e_n},\mathbb I-\op{e_n}{e_n}\}$.  From eq.~\eqref{eq:rhoea}, its relevant outcome $p_n=\bra{e_n}\rho_s^{(1)}\ket{e_n}=\Delta^{2n}
\left(|f_n|^2 x_{2n}+O(\Delta)\right)$, hence
\begin{align}
p_n=\Theta(\Delta^{2n}),\quad \partial_{x_{2n}}p_n=\Theta(\Delta^{2n}), \\
\Rightarrow
\mbb F_{2n,2n}^{(\mc M_{2n})}(\rho_s^{(1)})
\ge
\frac{(\partial_{x_{2n}}p_n)^2}{p_n}
=
\Theta(\Delta^{2n}).
\end{align}\par 
For $\mu=2n+1$, one can use the three-outcome measurement $\left\{\op{\phi_{n,\pm}}{\phi_{n,\pm}}, \mbb 1-\sum_{\pm}\op{\phi_{n,\pm}}{\phi_{n,\pm}}
\right\}$ with
\begin{align}
\ket{\phi_{n,\pm}}=\frac{\ket{e_n}\pm e^{i\chi_n}\ket{e_{n+1}}}{\sqrt2}.
\end{align}
The two probabilities $p_{n,\pm}=\bra{\phi_{n,\pm}}\rho_s^{(1)}\ket{\phi_{n,\pm}}$ satisfy
\begin{align}
p_{n,\pm}=\frac{1}{2}\Delta^{2n}\left[
|f_n|^2x_{2n}\pm 2\Delta\Re[f_nf^*_{n+1}e^{i\chi_n}]x_{2n+1}+O(\Delta)\right]\notag 
\end{align}
Hence $p_{n,\pm}=\Theta(\Delta^{2n})$, and
\begin{align}
\partial_{x_{2n+1}}p_{n,+}-\partial_{x_{2n+1}}p_{n,-}&=2\Delta^{2n+1}\Re[f_nf^*_{n+1}e^{i\chi_n}]\notag \\
&=\Theta(\Delta^{2n+1}).\notag
\end{align}
If $\chi_n$ is chosen so that the leading coefficient is real and nonzero, then
\begin{align}
&\mbb F_{2n+1,2n+1}^{(\mc M_{2n+1})}(\rho_s^{(1)})\ge \sum_{\sigma=\pm}\frac{(\partial_{x_{2n+1}}p_{n,\sigma})^2}{p_{n,\sigma}} \notag \\
&\ge\frac{
(\partial_{x_{2n+1}}p_{n,+}-\partial_{x_{2n+1}}p_{n,-})^2}{p_{n,+}+p_{n,-}}=\Theta(\Delta^{2n+2}).
\end{align}
Hence, with the upper bound in Eq.~\eqref{eq:app-qfi-upper} and the fact that $\mbb F_{\mu\mu}^{(\mc M_\mu)}(\rho^{(1)}_s)\le \mbb H_{\mu\mu}(\rho^{(1)}_s)$
\begin{align}
\mbb F_{\mu\mu}^{(\mc M_\mu)}(\rho^{(1)}_s)
=
\Theta(\Delta^{2\lceil\mu/2\rceil}),
\qquad
\mu\le2N-2.
\end{align}

Similarly, combining with Eq.~\eqref{eq:app-qfi-upper} proves
\begin{align}
\mbb H_{\mu\mu}(\rho_s^{(1)})
=
\begin{cases}
\Theta(\Delta^{2\lceil\mu/2\rceil}),
& \mu\le2N-2,\\
O(\Delta^{2\mu-2N+2}),
& \mu\ge2N-1.
\end{cases}
\end{align}
Finally, by the weak-source decomposition in Eq.~\eqref{eq:rho}, we have 
$\mbb H_{\mu\mu}(\rho_s)=\epsilon\mbb H_{\mu\mu}(\rho_s^{(1)})+O(\epsilon^2)$ and $\mbb F_{\mu\mu}(\rho_s)=\epsilon\mbb F_{\mu\mu}(\rho_s^{(1)})+O(\epsilon^2)$.

\paragraph*{Two projective settings.}
Now, we show that the moment-wise measurements above can be grouped into two projective measurements, since the block $\{\ket{\phi_{n,+}},\ket{\phi_{n,-}}\}$ not only
attains FI scaling of the odd moment $x_{2n+1}$, but also the even moment $x_{2n}$, since $\partial_{x_{2n}}p_{n,+}=\partial_{x_{2n}}p_{n,-}=\Theta(\Delta^{2n})$.

For even $N$, we can define two orthonormal bases by
\begin{align}
\mc B_{\mathrm{even}}&=\Bigl\{\ket{\phi_{2r,\pm}}:\ 2r\le N-2\Bigr\},\\
\mc B_{\mathrm{odd}}&=\{\ket{e_0},\ket{e_{N-1}}\}\cup
\Bigl\{\ket{\phi_{2r+1,\pm}}:\ 2r+1\le N-2\Bigr\}. \notag
\end{align}
For odd $N$, the same construction applies, except that the
endpoint $\ket{e_{N-1}}$ belongs to $\mc B_{\mathrm{even}}$ rather than
$\mc B_{\mathrm{odd}}$.  Hence, two projective settings attain the optimal FI scaling for all $\mu\le 2N-2$.
\section{Toy source example}
\label{app:toy-example}
Let $v=(u-u_0)/L\in[-1,1]$ and, for sufficiently small $|\eta|$, consider 
\begin{align}
I_\eta(u)=\frac{1}{L}p_\eta\left(\frac{u-u_0}{L}\right). \label{app:toy-source}
\end{align}
with $p_\eta(v)=\frac{1}{2}[1+\eta P_4(v)]$, and $P_4(v)=\frac{35v^4-30v^2+3}{8}$. By construction, the unknown parameter $\eta$ first appears through the fourth normalized moment, since 
\begin{align}
\left.\partial_\eta x_\mu\right|_{\eta=0}&=\frac{1}{2}\int_{-1}^1 dv v^\mu P_4(v)=0,\quad
\mu=0,1,2,3,\notag \\
\left.\partial_\eta x_4\right|_{\eta=0}&=\frac{1}{2}\int_{-1}^1 dv v^4 P_4(v)=\frac{8}{315}.
\label{eq:P4-x4-derivative}
\end{align}
Now, consider a three-telescope array with normalized coordinates $H=\mathrm{diag}(c_1,c_2,c_3)$ where 
$c_1=-1/2$, $c_2=0$, $c_3=1/2$ and $c_{ij}=c_i-c_j$.  One can define $\Delta=kLD/z_0$ as in Eq.~\eqref{eq:delta-def}. After absorbing the known point-source phases into the telescope modes, the one-photon source state is
\begin{align}
\rho_s^{(1)}(\eta,\Delta)=\frac{1}{3}\sum_{i,j=1}^3g_\eta(c_{ij}\Delta)\ket{1_i}\bra{1_j},
\label{eq:toy-one-photon-state}
\end{align}
where $g_\eta(a)=\int_{-1}^1dvp_\eta(v)e^{-iav}$. One can then obtain the array-SPADE vector via Eq~\eqref{GSbasis}, which yields
\begin{align}
\ket{e_0}&=\frac{\ket{1_1}+\ket{1_2}+\ket{1_3}}{\sqrt3}, \notag\\
\ket{e_1}&=\frac{-\ket{1_1}+\ket{1_3}}{\sqrt2},  \notag \\\ket{e_2}&=\frac{\ket{1_1}-2\ket{1_2}+\ket{1_3}}{\sqrt6}.
\label{eq:e2-three-telescope}
\end{align}
Projecting to $\ket{e_2}$ then yields outcome probability
\begin{align}
p_2(\eta,\Delta)&=\bra{e_2}\rho_s^{(1)}(\eta,\Delta)\ket{e_2}=\frac{3-4g_\eta(\Delta/2)+g_\eta(\Delta)}{9}. \notag 
\end{align}
At $\eta=0$, $g_{0}(a)=\frac{\sin a}{a}=1-\frac{a^2}{6}+\frac{a^4}{120}+O(a^6)$, and therefore
\begin{align*}
\left. p_2(\eta,\Delta)\right|_{\eta=0}&=\frac{\Delta^4}{1440}+O(\Delta^6)\\
\left.\partial_\eta p_2(\eta,\Delta)\right|_{\eta=0}&=\frac{\Delta^4}{11340}+O(\Delta^6).
\end{align*}
Consequently, using projective measurement $\mc M_4=\left\{\ket{e_2}\bra{e_2},\mathbb I-\ket{e_2}\bra{e_2}\right\}$, one has
\begin{align}
\left.F_{\eta\eta}^{(\mc M_4)}(\rho_s^{(1)})\right|_{\eta=0}
=\frac{2}{178605}\Delta^4+O(\Delta^6),
\end{align}
and hence $F_{\eta\eta}^{(\mc M_4)}(\rho_s)=\Theta(\epsilon\Delta^4)$.

By contrast, since $\eta$ has no first-order overlap with
$x_0,x_1,x_2$ and $x_3$, the pairwise measurement QFI bound gives $H_{\eta\eta}^{(\mathrm{pair})}=O(\epsilon\Delta^6)$ via Thm.~\ref{thm:pairwise-qfi-bound}.
Thus, for estimating $\eta$ in the toy source of Eq.~\eqref{app:toy-source}, genuinely three-telescope interference is necessary to attain the full-array
$\Theta(\epsilon\Delta^4)$ scaling; the projector onto $\ket{e_2}$ provides one
measurement that attains this scaling.

\onecolumngrid

\setcounter{section}{0}
\setcounter{equation}{0}
\setcounter{figure}{0}
\setcounter{table}{0}

\makeatletter
\@removefromreset{equation}{section}
\@removefromreset{figure}{section}
\@removefromreset{table}{section}
\makeatother

\renewcommand{\thesection}{S\arabic{section}}
\renewcommand{\theequation}{S\arabic{equation}}
\renewcommand{\thefigure}{S\arabic{figure}}
\renewcommand{\thetable}{S\arabic{table}}

\makeatletter
\renewcommand{\@seccntformat}[1]{\csname the#1\endcsname\quad}
\makeatother

\begin{center}
{\Large\bf Supplemental Material}
\end{center}
\section{Source reconstruction from image moments}
\label{app:moment-reconstruction}

In this appendix, we briefly clarify in what sense the image moments determine the source intensity distribution in the subdiffraction regime. Recall that the normalized image moments are
\begin{align}
x_\mu
=
\int du  I(u)\left(\frac{u-u_0}{L}\right)^\mu.
\label{eq:moment-reconstruction-def}
\end{align}
Here $u_0$ is a reference point on the source plane, and $I(u)$ is the normalized source intensity distribution, and $L$ characterizes the source size. 

Assume that the source is compact, so that $I(u)$ is supported on a bounded interval. Then the full infinite moment sequence $\{x_\mu\}_{\mu\ge 0}$ determines the source intensity distribution uniquely. Equivalently, define the characteristic function
\begin{align}
\Phi(t):=\int du  I(u)e^{-it(u-u_0)}=\sum_{\mu=0}^\infty
\frac{(-itL)^\mu}{\mu!} x_\mu,
\label{eq:char-fn-moment-expansion}
\end{align}
so the full moment sequence determines $\Phi(t)$, and hence $I(u)$ by inverse Fourier transform.

This uniqueness statement should be distinguished from finite-order reconstruction. If only the first $K$ moments are known, the source is generally not uniquely determined. Instead, the truncated moment constraints define a feasible family of nonnegative compactly supported intensities. A unique image then requires additional assumptions, such as a finite point-source model, sparsity, or a regularization principle such as maximum entropy~\cite{Narayan1986MaxEntAstronomy,Thiebaut2013Interferometry}.

In the subdiffraction regime, such a truncation is natural. Indeed, for the sampled spatial frequencies $t$ with $|tL|=|c_{jk}|\Delta\leq \Delta\ll 1$ in the subdiffraction regime, higher-order moments affect the sampled visibilities only through higher orders in $\Delta$ from Eq.~\eqref{eq:char-fn-moment-expansion}. Retaining only finitely many moments gives the natural low-frequency approximation to the sampled visibility expansion. In this sense, the image moments provide both a complete parametrization of a compact source when all orders are known, and a natural hierarchy of approximations in the subdiffraction limit when only finitely many moments are accessible.

We now derive the moment expansion of the one-photon source state defined in the main text
\begin{align}
(\rho_s^{(1)})_{jk}=\frac{1}{N}g_{jk}=\frac{1}{N}e^{i(\phi_j-\phi_k)}\int du I(u)\exp\left[-ic_{jk}\Delta\left(\frac{u-u_0}{L}\right)\right].
\end{align}
Expanding the expression above and using the definition of the normalized moments gives
\begin{align}
(\rho_s^{(1)})_{jk}&=\frac{1}{N}e^{i(\phi_j-\phi_k)}\sum_{\mu\ge 0}
\frac{(-i)^\mu}{\mu!}[c_{jk}\Delta]^\mu\int du I(u)\left(\frac{u-u_0}{L}\right)^\mu \notag \\
&=\frac{1}{N}e^{i(\phi_j-\phi_k)}\sum_{\mu\ge 0}\frac{(-i)^\mu}{\mu!}(c_{jk}\Delta)^\mu x_\mu=
\sum_{\mu\ge 0} x_\mu (A_\mu)_{jk},
\end{align}
where $c_{jk}=c_j-c_k$,  
\begin{align}
(A_\mu)_{jk}=&\frac{\Delta^\mu}{N}e^{i(\phi_j-\phi_k)}\frac{(-i)^\mu}{\mu!}
(c_j-c_k)^\mu=\frac{\Delta^\mu}{N}e^{i(\phi_j-\phi_k)}\sum_{m+n=\mu}
\frac{i^{n-m}}{m!n!}c_j^m c_k^n \notag\\
=&\left[\Delta^\mu\sum_{m+n=\mu}\frac{i^{n-m}}{m!n!}
H^m\ket{\psi}\bra{\psi}H^n
\right]_{jk}\label{eq:A-mu-matrix-elements},
\end{align}
with $\ket{\psi}=\frac{1}{\sqrt N}\sum_{j=1}^N e^{i\phi_j}\ket{1_j}_s$ and $H=\mathrm{diag}\left(c_1,\cdots, c_N\right)$.

\yk 
\section{Array-SPADE as a discrete aperture-plane SPADE}
\label{app:array-spade}

In this section, we clarify why the Gram--Schmidt basis
$\{\ket{e_a}\}_{a=0}^{N-1}$ used in the main text can be viewed as a
discrete version of the SPADE derivative-mode basis~\cite{Tsang2019,Tsang2021}. Simply put, in a single-aperture imaging system, the image-forming lens implements a spatial Fourier transform from aperture-plane modes to image-plane modes, where SPADE is usually defined. Pulling the SPADE derivative modes back to the aperture plane shows that they are generated by multiplying the reference aperture mode by powers of the aperture coordinate. Sampling this aperture-plane construction at the telescope positions gives the array-SPADE ladder used in the main text.

\subsection{Single-aperture SPADE and its aperture-plane representation}

To compare single-aperture SPADE, defined on the image plane, with array-SPADE, defined on the aperture plane of the telescope array, we first make explicit the spatial Quantum Fourier-transform unitary
\begin{align}
\mc F_{{\rm a}\to{\rm i}}\ket{c}_{\rm Ap}
\propto
\int dv \exp\left[i\frac{kD}{z_0}vc\right]\ket{v}_{\rm Img},
\label{eq:qft-a-i}
\end{align}
where $\ket{v}_{\rm Img}$ denotes an image-plane position basis state and
$\ket{c}_{\rm Ap}$ denotes an aperture-plane basis state. Here $c$ is a
dimensionless aperture coordinate, so that the physical aperture coordinate is
$Dc$. In a single-aperture imaging system, this spatial Fourier transform is
naturally implemented by the image-forming lens. Direct imaging then corresponds to
measuring the image-plane basis eigenstate $\{\ket{v}_{\rm Img}\}$, whereas SPADE measures an orthonormal spatial-mode basis represented in this image-plane basis.

For a uniform aperture $c\in[-1/2,1/2]$, a point source at
$u$ defines the aperture-plane one-photon state
\begin{align}
\ket{\widetilde \Psi(u)}_{\rm Ap}
=
\int_{-1/2}^{1/2}dc 
\exp\left[-i\frac{kD}{z_0}uc\right]\ket{c}_{\rm Ap}.
\end{align}
After the image-forming lens,
\begin{align}
\mc F_{{\rm a}\to{\rm i}}\ket{\widetilde\Psi(u)}_{\rm Ap}
&\propto
\int dv
\left[
\int_{-1/2}^{1/2}dc 
\exp\left[i\frac{kD}{z_0}(v-u)c\right]
\right]\ket{v}_{\rm Img}=:
\int dv \psi(v-u)\ket{v}_{\rm Img}
=
\ket{\Psi(u)}_{\rm Img},
\end{align}
where $\psi$ is the amplitude point-spread function. With our coordinate
convention, the image-plane displacement is written in source-coordinate units.
For a uniform one-dimensional aperture, $\psi$ is sinc-like.

We now recall the derivative-mode viewpoint on single-aperture SPADE in the
image plane. Taylor expanding $\ket{\Psi(u)}_{\rm Img}$ around $u_0$ gives
\begin{align}
\ket{\Psi(u)}_{\rm Img}
&=
\sum_{m\ge0}
\left(\frac{u-u_0}{L}\right)^m
\ket{\Psi^{(m)}}_{\rm Img},
\qquad
\ket{\Psi^{(m)}}_{\rm Img}
:=
\frac{1}{m!}
\left.
\left(L\frac{\partial}{\partial u}\right)^m
\ket{\Psi(u)}_{\rm Img}
\right|_{u=u_0}.
\label{eq:single-aperture-derivative}
\end{align}
For an incoherent source with normalized moments
$x_\mu=\int du I(u)[(u-u_0)/L]^\mu$, the one-photon state has the
derivative-mode expansion
\begin{align}
\rho
=
\int du I(u)\ket{\Psi(u)}\bra{\Psi(u)}
=
\sum_{m,n\ge0}
x_{m+n}\ket{\Psi^{(m)}}\bra{\Psi^{(n)}} .
\label{eq:single-aperture-t-expansion}
\end{align}
Thus the derivative-mode expansion has the same Hankel moment structure as the moment expansion used in the main text, and the SPADE basis is obtained by orthogonalizing the derivative-mode ladder
$\{\ket{\Psi^{(m)}}\}_{m=0}^{\infty}$.

Equivalently, one can define the same derivative-mode ladder in the
aperture-plane representation by pulling the image-plane modes back through
the inverse Fourier transform:
\begin{align}
\ket{\widetilde \Psi^{(0)}}_{\rm Ap}
:=
\mc F_{{\rm a}\to{\rm i}}^{-1}\ket{\Psi^{(0)}}_{\rm Img}
\propto
\int_{-1/2}^{1/2}dc 
\exp\left[-i\frac{kD}{z_0}u_0c\right]\ket{c}_{\rm Ap}.
\label{eq:aperture-plane-point}
\end{align}
Similarly,
\begin{align}
\ket{\widetilde \Psi^{(m)}}_{\rm Ap}
:=
\mc F_{{\rm a}\to{\rm i}}^{-1}\ket{\Psi^{(m)}}_{\rm Img}
\propto
\int_{-1/2}^{1/2}dc c^m
\exp\left[-i\frac{kD}{z_0}u_0c\right]\ket{c}_{\rm Ap},
\label{eq:aperture-plane-derivative}
\end{align}
Thus, in the aperture representation, the SPADE
derivative-mode ladder is obtained by multiplying the reference aperture mode by powers of $c$. And the SPADE basis is then obtained by orthogonalizing the derivative-mode ladder $\{\ket{\widetilde\Psi^{(m)}}\}_{m=0}^{\infty}$.

One interpretation of the result is that the SPADE measurement defined on the aperture plane can be decomposed into a Quantum Fourier transform followed by a SPADE measurement on the image plane.

\subsection{Array-SPADE as discrete aperture-plane SPADE}

We now pass to the telescope array, where the continuous aperture coordinate $c$ is
sampled only at the telescope positions $c_j=d_j/D$, with
$D=\max_{j,k}|d_j-d_k|$. Defining
\begin{align}
\phi_j
:=
-\frac{k u_0D c_j}{z_0},
\qquad
\ket{\psi}_{\rm Ap}
=
\frac{1}{\sqrt N}
\sum_{j=1}^N e^{i\phi_j}\ket{c_j}_{\rm Ap},
\qquad
H=\mathrm{diag}(c_1,\ldots,c_N),
\label{eq:psi-array-SI}
\end{align}
the sampled version of Eq.~\eqref{eq:aperture-plane-derivative} becomes, up to
the scalar prefactor,
\begin{align}
\ket{\widetilde\Psi^{(m)}}_{\rm Ap}
:=
H^m\ket{\psi}_{\rm Ap}
=
\frac{1}{\sqrt N}
\sum_{j=1}^N c_j^m e^{i\phi_j}\ket{c_j}_{\rm Ap}.
\label{eq:sampled-derivative-mode}
\end{align}
Here the equal weights follow from sampling the uniform aperture envelope. More
general known telescope weights would simply replace $\ket{\psi}_{\rm Ap}$ by
the corresponding weight.

Thus the state vectors
$\{\ket{\widetilde\Psi^{(0)}}_{\rm Ap},
\ket{\widetilde\Psi^{(1)}}_{\rm Ap},
\ket{\widetilde\Psi^{(2)}}_{\rm Ap},\ldots\}$ are the discretized
aperture-plane versions of the single-aperture SPADE derivative-mode ladder.
Consequently,
\begin{align}
\mc K
=
\mathrm{Span}\{\ket{\psi}_{\rm Ap},
H\ket{\psi}_{\rm Ap},
H^2\ket{\psi}_{\rm Ap},\ldots\}
\end{align}
defines the discrete SPADE basis associated with the synthesized aperture of
the telescope array.

This is the sense in which array-SPADE can be regarded as a Quantum Fourier
transformation followed by a discrete single-aperture SPADE measurement. In
practice, the two stages need not appear as separate optical elements; they can
be compiled into a single multi-telescope interferometric measurement.

This interpretation also clarifies the distinction from
Refs.~\cite{Sajjad2024,Padilla2026,Padilla2026a}. In our approach, the array-SPADE measurement is spread over all telescopes, and inter-telescope interference
implements the mode projection itself. In
contrast, Refs.~\cite{Sajjad2024,Padilla2026,Padilla2026a} perform SPADE locally at each telescope and then interfere the corresponding sorted spatial modes pairwise. 
\blk
\section{Pairwise interferometric protocols}
\subsection{An improved pairwise quantum Fisher information bound}
\label{app:improved-pairwise-qfi}

\begin{lemma}
\label{cor:pairwise-qfi-sharp}
Under the source \textit{nondegeneracy} condition used in the main text,
the pairwise benchmark for estimating the normalized image moment $x_\mu$
satisfies
\begin{align}
\mbb{H}^{\mathrm{(pair)}}_{\mu\mu}(\rho_s)
=
\begin{cases}
\Theta\bigl(\epsilon\Delta^{2\mu}\bigr), & \mu~~\mathrm{odd},\\
\Theta\bigl(\epsilon\Delta^{2\mu-2}\bigr), & \mu~~\mathrm{even}.
\end{cases}
\label{eq:pairwise-qfi-sharp}
\end{align}
\end{lemma}

Fix a telescope pair $(i,j)$ with $c_{ij}\neq0$. After absorbing the known
point-source phase into the mode definition, the normalized one-photon
pair-restricted state is
\begin{align}
\rho_{ij}^{(1)}
:=
\frac{P_{ij}\rho_s^{(1)}P_{ij}}
{\tr(P_{ij}\rho_s^{(1)})}
=
\frac12
\begin{pmatrix}
1 & g(c_{ij}\Delta)\\
g(c_{ij}\Delta)^* & 1
\end{pmatrix},
\qquad
g(c_{ij}\Delta)
=
\sum_{\nu\ge0}\frac{(-ic_{ij}\Delta)^\nu}{\nu!}x_\nu .
\label{eq:pair-state-g}
\end{align}
For a two-dimensional state of this form, the QFI with respect to $x_\mu$ is
\begin{align}
\mbb H_{\mu\mu}\left(\rho_{ij}^{(1)}\right)
=
\left|\partial_{x_\mu}g(c_{ij}\Delta)\right|^2
+
\frac{
\left[
\text{Re}\left(
g(c_{ij}\Delta)^*
\partial_{x_\mu}g(c_{ij}\Delta)
\right)
\right]^2
}
{1-\left|g(c_{ij}\Delta)\right|^2}.
\label{eq:two-mode-qfi}
\end{align}
Since $\partial_{x_\mu}g(c_{ij}\Delta)=\frac{(-ic_{ij}\Delta)^\mu}{\mu!}$, one has $
\left|\partial_{x_\mu}g(c_{ij}\Delta)\right|^2=\Theta(\Delta^{2\mu})$. Moreover $1-\left|g(c_{ij}\Delta)\right|^2=
(x_2-x_1^2)(c_{ij}\Delta)^2+O(\Delta^4)$, the strict positivity of the $2\times2$ principal minor of the Hankel matrix $\Gamma$ implied by the nondegeneracy assumption gives $x_2-x_1^2>0$ (i.e., $\Gamma_{\nu\mu}=x_{\nu+\mu}$.),  therefore, one has $1-\left|g(c_{ij}\Delta)\right|^2=\Theta(\Delta^2)$.

It remains to determine the parity-dependent scaling of
\begin{align}
\text{Re}\left[
g(c_{ij}\Delta)^*\partial_{x_\mu}g(c_{ij}\Delta)\right]=\text{Re}\left[
g(c_{ij}\Delta)^*
\frac{(-ic_{ij}\Delta)^\mu}{\mu!}
\right].
\label{eq:real-term}
\end{align}

\paragraph*{Even moments.} The $\nu=0$ contribution from
$g(c_{ij}\Delta)^*$ gives a real nonzero leading term in
Eq.~\eqref{eq:pair-state-g}. Hence
Eq.~\eqref{eq:real-term} is $\Theta(\Delta^\mu)$, and the second term in
Eq.~\eqref{eq:two-mode-qfi} thus scales as $\Theta(\Delta^{2\mu-2})$, dominating
the first term $\Theta(\Delta^{2\mu})$. Thus
\begin{align}
\mbb H_{\mu\mu}\left(\rho_{ij}^{(1)}\right)
=
\Theta(\Delta^{2\mu-2}),
\qquad
\mu \ \mathrm{even}.
\label{eq:pair-even-qfi}
\end{align}
\paragraph*{Odd moments.}
The $\nu=0$ contribution is purely imaginary and therefore
vanishes under the real part. The first possible real contribution is one order
higher, i.e., with $\nu=1$ term from
$g(c_{ij}\Delta)^*$. Thus Eq.~\eqref{eq:real-term} is $O(\Delta^{\mu+1})$, so the second term in
Eq.~\eqref{eq:two-mode-qfi} is $O(\Delta^{2\mu})$. Together with the first term
$\Theta(\Delta^{2\mu})$, this gives
\begin{align}
\mbb H_{\mu\mu}\left(\rho_{ij}^{(1)}\right)
=
\Theta(\Delta^{2\mu}),
\qquad
\mu\ \mathrm{odd}.
\label{eq:pair-odd-qfi}
\end{align}
Finally, the normalized pair-restricted one-photon state occurs in the original
weak-light state with weight $\tr(P_{ij}\rho_sP_{ij})=\frac{2\epsilon}{N}+O(\epsilon^2)$, hence the scaling of $\mbb{H}^{\mathrm{(pair)}}_{\mu\mu}(\rho_s)$ is given in Eq~\eqref{eq:pairwise-qfi-sharp}.

\subsection{Standard pairwise heterodyne measurement}
\label{app:heterodyne}
Gaussian measurements are known to be insufficient for attaining the quantum superresolution scalings of weak thermal-light imaging in general~\cite{Wang2025}. This includes both local and nonlocal Gaussian measurements, even when arbitrary noiseless beam-splitter operations between light received at different telescopes are allowed. In the following, we explore several typical Gaussian measurements for imaging subdiffraction sources. As we will see, none of the Gaussian measurements considered here can achieve the preferred optimal Fisher-information scaling.

In this subsection, we derive the Fisher information scaling for local heterodyne measurement on a fixed telescope pair $(i,j)$ in an $N$-telescope array. Let $\alpha_i,\alpha_j\in\mathbb C$ denote the heterodyne outcomes on modes $i$ and $j$. The corresponding POVM (Here and below, the two-mode POVM is understood to act as identity on all unmeasured modes) is
\begin{align}
\Pi(\alpha_i,\alpha_j)
=
\frac{1}{\pi^2}
\ket{\alpha_i,\alpha_j}\bra{\alpha_i,\alpha_j}.
\end{align}
Using the weak-light expansion in Eq.~(5) of the main text, one obtains
\begin{align}
&p^{\rm het}_{ij}(\alpha_i,\alpha_j)=
\tr\bigl[\rho_{s}\Pi(\alpha_i,\alpha_j)\bigr]=\frac{e^{-|\alpha_i|^2-|\alpha_j|^2}}{\pi^2}\left[1+\frac{\epsilon}{N}\Bigl(|\alpha_i|^2+|\alpha_j|^2-2
+g_{ij}\alpha_i^*\alpha_j+g_{ij}^*\alpha_i\alpha_j^*\Bigr)\right]+O(\epsilon^2).
\label{eq:phet}
\end{align}
Hence
\begin{align}
\partial_{x_\mu}p^{\rm het}_{ij}=\frac{2\epsilon}{N}\frac{e^{-|\alpha_i|^2-|\alpha_j|^2}}{\pi^2}\Re\left[(\partial_{x_\mu}g_{ij})\alpha_i^*\alpha_j\right]
+O(\epsilon^2).
\label{eq:dphet}
\end{align}
To leading order in $\epsilon$, the denominator may be replaced by $p_0^{\rm het}(\alpha_i,\alpha_j)
=\frac{e^{-|\alpha_i|^2-|\alpha_j|^2}}{\pi^2}.$
The Fisher information is therefore
\begin{align}
&F^{\rm het}_{ij,\mu\mu}=\iint d^2\alpha_id^2\alpha_j\frac{\bigl(\partial_{x_\mu}p^{\rm het}_{ij}\bigr)^2}
{p^{\rm het}_{ij}}=\frac{4\epsilon^2}{N^2}\iint d^2\alpha_id^2\alpha_jp_0(\alpha_i,\alpha_j)
\left[\Re\left((\partial_{x_\mu}g_{ij})\alpha_i^*\alpha_j\right)\right]^2+O(\epsilon^3).\notag\\
&=\frac{2\epsilon^2}{N^2}\left|\partial_{x_\mu}g_{ij}\right|^2+O(\epsilon^3), \label{eq:Fhet-pre}
\end{align}
where $\iint d^2\alpha_id^2\alpha_j
p_0(\alpha_i,\alpha_j)
\left[
\Re\left(t\alpha_i^*\alpha_j\right)
\right]^2
=
\frac{|t|^2}{2}$. Since $\partial_{x_\mu}g_{ij}
=
e^{i(\phi_{ij})}
\frac{(-ic_{ij}\Delta)^\mu}{\mu!}$,  substituting into Eq.~\eqref{eq:Fhet-pre} gives
\begin{align}
F^{\rm het}_{ij,\mu\mu}
=
\Theta(\epsilon^2\Delta^{2\mu}),
\label{eq:Fhet-fixedpair}
\end{align}
for any fixed pair with $c_{ij}\neq 0$. Maximizing over telescope pairs does not change the $\Delta$-scaling. Thus, the standard pairwise heterodyne measurement satisfies
\begin{align}
F^{\rm het}_{\mu\mu}(\rho_s)
=
\Theta(\epsilon^2\Delta^{2\mu}).
\label{eq:Fhet-final}
\end{align}

Equation~\eqref{eq:Fhet-final} shows that, although heterodyne accesses the full complex visibility $g_{ij}$ rather than only $|g_{ij}|^2$, it remains a local Gaussian measurement on weak thermal light and therefore suffers an $\epsilon^2$ scaling~\cite{Wang2025}. Moreover, unlike the pairwise coherent scheme in Theorem~2 in the main text, it does not exploit the pair-restricted one-photon structure sufficiently to achieve the improved $\Delta$-scaling for low-order moments as in the optimal pairwise scheme.

\subsection{Standard pairwise homodyne measurement}
\label{app:homodyne}

In this subsection, we derive the Fisher information scaling for local homodyne measurement on a fixed telescope pair $(i,j)$ in an $N$-telescope array. 

For local-oscillator phases $\theta_i$ and $\theta_j$, let $y_i,y_j\in\mathbb R$ denote the corresponding quadrature outcomes, and let $\ket{y_i;\theta_i}$ and $\ket{y_j;\theta_j}$ be the quadrature eigenstates. The homodyne POVM that acts on the telescope pair $(i,j)$  is
\begin{align}
\Pi_{\theta_i,\theta_j}(y_i,y_j)
=
\ket{y_i;\theta_i}\bra{y_i;\theta_i}
\otimes
\ket{y_j;\theta_j}\bra{y_j;\theta_j},
\end{align}
with normalization $\iint dy_idy_j
\Pi_{\theta_i,\theta_j}(y_i,y_j)
=
\mathbb{1}$. Using the weak-light expansion in Eq.~(5) in the main , one obtains the joint homodyne outcome density
\begin{align}
p^{\rm hom}_{ij}(y_i,y_j)=
\tr\left[
\rho_{s}\Pi_{\theta_i,\theta_j}(y_i,y_j)
\right]=
\frac{e^{-y_i^2-y_j^2}}{\pi}
\left[
1+\frac{\epsilon}{N}
\Bigl(
2y_i^2+2y_j^2-2
+
4y_iy_j
\Re\left(
e^{-i(\theta_i-\theta_j)}g_{ij}
\right)
\Bigr)
\right]
+O(\epsilon^2).
\label{eq:phom}
\end{align}
Hence
\begin{align}
\partial_{x_\mu}p^{\rm hom}_{ij}
=
\frac{4\epsilon}{N}
\frac{e^{-y_i^2-y_j^2}}{\pi}
y_iy_j
\Re\left[
e^{-i(\theta_i-\theta_j)}
\partial_{x_\mu}g_{ij}
\right]
+O(\epsilon^2).
\label{eq:dphom}
\end{align}
To leading order in $\epsilon$, the denominator in the Fisher information may be replaced by $p^{\rm hom}_0(y_i,y_j)
=
\frac{e^{-y_i^2-y_j^2}}{\pi}$, therefore,
\begin{align}
&F^{\rm hom}_{ij,\mu\mu}
=
\iint dy_idy_j
\frac{\bigl(\partial_{x_\mu}p^{\rm hom}_{ij}\bigr)^2}
{p^{\rm hom}_{ij}}=
\frac{16\epsilon^2}{N^2}
\iint dy_idy_j
p^{\rm hom}_0(y_i,y_j)y_i^2y_j^2
\left[
\Re\left(
e^{-i(\theta_i-\theta_j)}
\partial_{x_\mu}g_{ij}
\right)
\right]^2+O(\epsilon^3)
\notag\\
&=
\frac{4\epsilon^2}{N^2}
\left[
\Re\left(
e^{-i(\theta_i-\theta_j)}
\partial_{x_\mu}g_{ij}
\right)
\right]^2+O(\epsilon^3),
\label{eq:Fhom-pre}
\end{align}
where $\iint dy_idy_j p^{\rm hom}_0(y_i,y_j)y_i^2y_j^2=\frac{1}{4}$. Optimizing over the local-oscillator phase difference $\theta_i-\theta_j$ gives, one thus has
\begin{align}
F^{\rm hom}_{ij,\mu\mu}
=
\frac{4\epsilon^2}{N^2}
\left|
\partial_{x_\mu}g_{ij}
\right|^2
+O(\epsilon^3).
\label{eq:Fhom-g}
\end{align}
Since $\partial_{x_\mu}g_{ij}
=
e^{i(\phi_{ij})}
\frac{(-ic_{ij}\Delta)^\mu}{\mu!}$, substitution into Eq.~\eqref{eq:Fhom-g} gives
\begin{align}
F^{\rm hom}_{ij,\mu\mu}
=
\Theta(\epsilon^2\Delta^{2\mu}),
\label{eq:Fhom-fixedpair}
\end{align}
for any fixed pair with $c_{ij}\neq 0$. Maximizing over telescope pairs does not change the $\Delta$-scaling. Thus the standard pairwise homodyne measurement satisfies
\begin{align}
F^{\rm hom}_{\mu\mu}(\rho_s)
=
\Theta(\epsilon^2\Delta^{2\mu}).
\label{eq:Fhom-final}
\end{align}

\subsection{Standard pairwise HBT measurement}
\label{app:HBT}

In this subsection, we derive the Fisher information scaling for the two-detector Hanbury Brown--Twiss (HBT) measurement (intensity interferometer) on a fixed telescope pair $(i,j)$ in an $N$-telescope array.

Let $n_k:=a_k^\dagger a_k$ denote the photon-number operator at telescope $k$. The standard HBT record is the binary coincidence/no-coincidence outcome. For the weak thermal light (a Gaussian state) considered in the main text, the coincidence probability agrees to leading order with the normally ordered second-order moment
\begin{align}
\langle n_i n_j\rangle=\langle n_i\rangle \langle n_j\rangle+\bigl|\langle a_i^\dagger a_j\rangle\bigr|^2,
\qquad i\neq j.
\label{eq:HBT-Wick}
\end{align}
Using the weak-light expansion in Eq.~(5) in the main , one has
\begin{align}
\langle n_i\rangle &= \frac{\epsilon}{N}+O(\epsilon^2), \\
\langle a_i^\dagger a_j\rangle &= \frac{\epsilon}{N} g_{ij}+O(\epsilon^2).
\end{align}
Therefore, coincidence probability on the pair $(i,j)$ is
\begin{align}
p^{\rm HBT}_{ij}
:=
\langle n_i n_j\rangle
=
\frac{\epsilon^2}{N^2}\bigl(1+|g_{ij}|^2\bigr)+O(\epsilon^3).
\label{eq:HBT-coincidence}
\end{align}

Since $p^{\rm HBT}_{ij}=\Theta(\epsilon^2)$, treating the HBT measurement as a binary coincidence/no-coincidence measurement, the Fisher information for estimating the moment parameter $x_\mu$ is
\begin{align}
F^{\rm HBT}_{ij,\mu\mu}
&=
\frac{\bigl(\partial_{x_\mu}p^{\rm HBT}_{ij}\bigr)^2}
{p^{\rm HBT}_{ij}\bigl(1-p^{\rm HBT}_{ij}\bigr)}
=
\frac{\epsilon^2}{N^2}
\frac{\bigl(\partial_{x_\mu}|g_{ij}|^2\bigr)^2}
{1+|g_{ij}|^2}
+O(\epsilon^3).
\label{eq:HBT-FI-prelim}
\end{align}
Thus, the problem reduces to the scaling of $\partial_{x_\mu}|g_{ij}|^2$.

Using Eq.~(4) of the main text, one may expand $g_{ij}=e^{i\phi_{ij}}\sum_{\nu\ge 0}\frac{(-ic_{ij}\Delta)^\nu}{\nu!}x_\nu$. Therefore, 
\begin{align}
\partial_{x_\mu}|g_{ij}|^2=2\Re\left[\frac{(-ic_{ij}\Delta)^\mu}{\mu!}
e^{i\phi_{ij}}g_{ij}^*\right]=2\Re\left[\frac{(-ic_{ij}\Delta)^\mu}{\mu!}
\sum_{\nu\ge 0}\frac{(ic_{ij}\Delta)^\nu}{\nu!}x_\nu\right].
\label{eq:dmodg}
\end{align}

\paragraph*{Even moments.}
Let $\mu=2n$. Since $(-i)^{2n}=(-1)^n$ is real, the leading contribution in Eq.~\eqref{eq:dmodg} comes from the $\nu=0$ term:
\begin{align}
\partial_{x_{2n}}|g_{ij}|^2
=
2(-1)^n\frac{(c_{ij}\Delta)^{2n}}{(2n)!}
+O(\Delta^{2n+2})
=
\Theta(\Delta^{2n}),
\label{eq:even-derivative-HBT}
\end{align}
provided $c_{ij}\neq 0$. Substituting into Eq.~\eqref{eq:HBT-FI-prelim} yields
\begin{align}
F^{\rm HBT}_{ij,2n,2n}
=
\Theta(\epsilon^2\Delta^{4n})
=
\Theta(\epsilon^2\Delta^{2\mu}).
\label{eq:even-FI-HBT}
\end{align}

\paragraph*{Odd moments.}
Let $\mu=2n+1$. Then $(-i)^{2n+1}$ is purely imaginary, so the $\nu=0$ term in Eq.~\eqref{eq:dmodg} does not contribute. The first potentially nonzero contribution comes from the $\nu=1$ term, namely
\begin{align}
\partial_{x_{2n+1}}|g_{ij}|^2
=
2(-1)^n\frac{x_1(c_{ij}\Delta)^{2n+2}}{(2n+1)!}
+O(\Delta^{2n+4}).
\label{eq:odd-derivative-HBT}
\end{align}
Hence, 
\begin{align}
F^{\rm HBT}_{ij,2n+1,2n+1}
=
O(\epsilon^2\Delta^{4n+4})
=
O(\epsilon^2\Delta^{2\mu+2}),
\label{eq:odd-FI-HBT}
\end{align}
since $x_1\ne 0$ is not assumed in our discussion (i.e., $\Gamma>0$ does not imply it). 

Combining Eqs.~\eqref{eq:even-FI-HBT} and \eqref{eq:odd-FI-HBT}, we obtain the Fisher information scaling for the standard pairwise HBT (i.e., intensity interferometer) scheme:
\begin{align}
F^{\rm HBT}_{ij,\mu\mu}
=
\begin{cases}
\Theta(\epsilon^2\Delta^{2\mu}), & \mu~~\mathrm{ even},\\[4pt]
O(\epsilon^2\Delta^{2\mu+2}), & \mu~~\mathrm{ odd }.
\end{cases}
\label{eq:HBT-fixed-pair}
\end{align}

Equation~\eqref{eq:HBT-fixed-pair} shows explicitly why standard HBT performs much worse than coherent pairwise interferometry for moment estimation in the subdiffraction regime: the coincidence signal depends on $|g_{ij}|^2$ rather than the full complex visibility $g_{ij}$, and is therefore both $\epsilon$-suppressed and less sensitive to the moment-dependent perturbations of $g_{ij}$. This obstruction is specific to the standard pairwise HBT measurement considered here; genuinely multi-telescope intensity-correlation measurements may access image-moment information differently.

\section{Memory-based implementation: write, erase, parity check, and measure}
\label{app:memory-assist}

In this appendix, we give a proof-of-principle memory-based implementation of the genuine multi-telescope measurements sketched in the main text. The goal is not to optimize the resource cost, but to show that, conditioned on heralded success events, the distributed one-photon source qudit can be coherently encoded into a distributed memory qudit without revealing the excitation location. Once this encoding is available, rank-one effects on the one-photon source subspace can be implemented by measuring the corresponding projectors on the memory qudit.

We keep the vacuum subspace explicitly because the coherent-state erasure step is postselected and can leave a residual vacuum branch in the memory register. Accordingly, we work on the vacuum-plus-one-excitation subspace
\begin{align}
\mc H_t^{(0,1)}
=
\mathrm{Span}\{
\ket{\mathrm{vac}}_t,\ket{1_1}_t,\dots,\ket{1_N}_t
\},
\quad
t\in\{s,m,h\},
\end{align}
where $\ket{\mathrm{vac}}_t=\ket{0\cdots0}_t$ and $\ket{1_j}_t=\ket{0\cdots1_j\cdots0}_t$ with $s,m,h$ denoting the source, memory, and herald registers, respectively.

\paragraph*{Write.}
The idealized write step can be described as the isometry
\begin{align}
V_{\rm w}
=
\ket{\mathrm{vac}}_m\ket{\mathrm{vac}}_h\bra{\mathrm{vac}}_s
+
\sum_{j=1}^N
\ket{1_j}_m\ket{1_j}_h\bra{1_j}_s.
\label{eq:Vw}
\end{align}
For illustration, let this isometry act on a pure vector in this subspace:
\begin{align}
\ket{\psi}_s=\psi_0\ket{\mathrm{vac}}_s+\sum_{j=1}^N \psi_j\ket{1_j}_s,
\label{eq:psi-in}
\end{align}
which gives
\begin{align}
\ket{\Psi_{\rm w}}_{mh}=V_{\rm w}\ket{\psi}_s
=\psi_0\ket{\mathrm{vac}}_m\ket{\mathrm{vac}}_h+\sum_{j=1}^N \psi_j\ket{1_j}_m\ket{1_j}_h.
\label{eq:psi-write}
\end{align}

\paragraph*{Erase.}
We model the erasure step as an $N$-site generalization of coherent-state photon-mode erasure used in memory-assisted nonlocal interferometry~\cite{Stas2026}. At each telescope $j$, the herald mode $h_j$ is mixed on a 50:50 beam splitter with a local coherent state $\ket{\alpha}_{l_j}$ at local oscillator mode $l_j$; the local oscillators are assumed to have equal intensity and a shared phase reference. The two output ports are then measured with photon-number-resolving detectors. Let $(n_j^+,n_j^-)\in\mathbb N^2$ denote the detected counts at site $j$.

For a fixed local outcome $(p,q)$ with $p$ and $q$ being the photon number detected at the two outputs, the induced measurement on the herald mode can be represented as:
\begin{align}
\bra{\xi_j^{(p,q)}}
:=
\bra{p,q}U_{{\rm BS},j}\ket{\alpha}_{l_j}
=
A^{(p,q)}\bra{0}_{h_j}
+
B^{(p,q)}\bra{1}_{h_j},
\label{eq:local-bra}
\end{align}
with
\begin{subequations}
\begin{align}
A^{(p,q)}
&=
e^{-|\alpha|^2/2}
\frac{(-1)^q}{\sqrt{p!q!}}
\left(\frac{\alpha}{\sqrt2}\right)^{p+q},
\quad\quad B^{(p,q)}=
e^{-|\alpha|^2/2}
\frac{(-1)^q(p-q)}{\sqrt{2p!q!}}
\left(\frac{\alpha}{\sqrt2}\right)^{p+q-1}.
\end{align}
\end{subequations}
Thus $B^{(p,q)}/A^{(p,q)}=(p-q)/\alpha$ for $\alpha\neq0$. The dependence on arbitrary photon counts $(p,q)$ is what accounts for the vacuum, single-photon, and higher-photon-number components of the coherent erasure field.

For a global click pattern $\mbf n=\{(n_1^+,n_1^-),\dots,(n_N^+,n_N^-)\}$, the erase step, together with the write isometry, induces the following Kraus operator from the source subspace to the memory subspace:
\begin{align}
K_{\mbf n}
&:=
\left(
\bigotimes_{j=1}^N
\bra{\xi_j^{(n_j^+,n_j^-)}}
\right)V_{\rm w}=
\left(\prod_{j=1}^N A^{(n_j^+,n_j^-)}\right)
\ket{\mathrm{vac}}_m\bra{\mathrm{vac}}_s+
\sum_{j=1}^N
B^{(n_j^+,n_j^-)}
\prod_{k\neq j}
A^{(n_k^+,n_k^-)}
\ket{1_j}_m\bra{1_j}_s .
\label{eq:erase-kraus}
\end{align}
Following strategy 1 of the Photon erasure step in Ref.~\cite{Stas2026}, if one accepts only click patterns satisfying $|n_j^+-n_j^-|=r$ $(r\ge1)$, denote the accepted pattern by $\mbf n_r$ and define $s_j:=\mathrm{sgn}(n_j^+-n_j^-)$, and $\Gamma_{\mbf n_r}:=
\prod_{j=1}^N A^{(n_j^+,n_j^-)}$. Then Eq.~\eqref{eq:erase-kraus} becomes
\begin{align}
K_{\mbf n_r}
=
\Gamma_{\mbf n_r}
\left[
\ket{\mathrm{vac}}_m\bra{\mathrm{vac}}_s
+
\frac{r}{\alpha}
\sum_{j=1}^N
s_j\ket{1_j}_m\bra{1_j}_s
\right].
\label{eq:accepted-kraus}
\end{align}
The outcome-dependent signs can be removed by local feedforward gate $F_{\mbf s}=\prod_{j=2}^NZ_{m_j}^{(1-s_1s_j)/2}$, which gives
\begin{align}
F_{\mbf s}K_{\mbf n_r}
=
\Gamma_{\mbf n_r}
\left[
\ket{\mathrm{vac}}_m\bra{\mathrm{vac}}_s
+
\frac{r}{\alpha}s_1
\sum_{j=1}^N
\ket{1_j}_m\bra{1_j}_s
\right].
\label{eq:feedforward-kraus}
\end{align}
For the illustrative input state in Eq.~\eqref{eq:psi-in}, this gives
\begin{align}
&\ket{\Psi_{\mbf n_r}}_m= F_{\mbf s}K_{\mbf n_r}\ket{\psi}_s=\Gamma_{\mbf n_r}
\left(\psi_0\ket{\mathrm{vac}}_m+\frac{r}{\alpha} s_1\sum_{j=1}^N \psi_j\ket{1_j}_m
\right).\label{eq:psi-after-erase}
\end{align}
Thus, up to an overall amplitude, every accepted branch transfers the source amplitudes coherently to the memory register, but with a residual vacuum component. Different values of $r$ lead to different vacuum-to-signal ratios.

\paragraph*{Parity check.}
The residual vacuum branch in Eq.~\eqref{eq:feedforward-kraus} can be removed without revealing the excitation location by measuring $S=\prod_{j=1}^N Z_{m_j}$. Since $S\ket{\mathrm{vac}}_m=\ket{\mathrm{vac}}_m$, and $S\ket{1_j}_m=-\ket{1_j}_m,$ for all $j$, the odd-parity projector $\Pi_{\rm odd}=\frac{\mathbb I-S}{2}$  annihilates the vacuum state and acts as the identity on the single-excitation subspace $\mc H_m^{(1)}$. 

Generalizing the 2-qubit parity check construction in~\cite{Khabiboulline2019PRA,Stas2026}, a standard distributed implementation of this parity measurement $\{\frac{\mathbb I-S}{2}, \frac{\mathbb I+S}{2}\}$ uses:

\begin{itemize}
    \item An ancillary $N$-qubit GHZ state: $\ket{\mathrm{GHZ}}_a= \frac{\ket{0\cdots 0}_a+\ket{1\cdots 1}_a}{\sqrt2}$

    \item Local controlled-$Z$ gates between $a_j$ and $m_j$:  $\prod_{j=1}^N CZ_{a_jm_j}$ 

    \item Projective measurement of the ancillas onto the $X$-basis: $\ket{\mbf X}_a=\bigotimes_{j=1}^N \frac{\ket{0}_a+\eta_j\ket{1}_a}{\sqrt2}
$ for $\eta_j\in\{\pm1\}$.
\end{itemize}

These together induce the following operator on the memories:
\begin{align}
&(\bra{\mbf X}_a\otimes \mathbb I)
\left(\prod_{j=1}^N CZ_{a_jm_j}\right)
\ket{\mathrm{GHZ}}_a= 2^{-(N+1)/2}
\left(\mathbb I + \Bigl(\prod_{j=1}^N \eta_j\Bigr) S\right).
\label{eq:GHZ-parity}
\end{align}
Hence coarse graining all outcomes with $\prod_j \eta_j=-1$ implements the odd-parity projector $\Pi_{\rm odd}=\frac{\mathbb I-S}{2}$ up to normalization. Applying $\Pi_{\rm odd}$ to Eq.~\eqref{eq:feedforward-kraus} gives
\begin{align}
\Pi_{\rm odd}F_{\mbf s}K_{\mbf n_r}
=
\Gamma_{\mbf n_r}
\frac{r}{\alpha}s_1
\sum_{j=1}^N
\ket{1_j}_m\bra{1_j}_s .
\label{eq:heralded-isometry}
\end{align}
For example, take the state in Eq.~\eqref{eq:psi-in} as an input. One finds
\begin{align}
&\ket{\Psi_{\mbf n_r}}_m= \Pi_{\rm odd}F_{\mbf s}K_{\mbf n_r}\ket{\psi}_s=\Gamma_{\mbf n_r}
\frac{r}{\alpha} s_1\sum_{j=1}^N \psi_j\ket{1_j}_m. \label{eq:psi-after-parity}
\end{align}

For an arbitrary input $\rho_s$ supported on $\mc H_s^{(0,1)}$, the success probability of a particular accepted branch is
\begin{align}
p(\mbf n_r|\rho_s)
&=\tr\left[
(\Pi_{\rm odd}F_{\mbf s}K_{\mbf n_r})
\rho_s
(\Pi_{\rm odd}F_{\mbf s}K_{\mbf n_r})^\dagger
\right]=
|\Gamma_{\mbf n_r}|^2
\left|\frac{r}{\alpha}\right|^2
\tr\left[
\Pi_s^{(1)}\rho_s
\right],
\label{eq:branch-success-general}
\end{align}
where $\Pi_s^{(1)}$ is the projector onto the source one-photon subspace. For the weak-light state in Eq.~(5) of the main text, the corresponding unconditional probability is multiplied by the one-photon weight, $\epsilon+O(\epsilon^2)$. In what follows, we quote the success probability conditioned on the source being in the one-photon subspace, i.e., with $\tr[\Pi_s^{(1)}\rho_s]=1$.

Summing over all accepted patterns $\mbf n_r$ with fixed $r$ gives
\begin{align}
P_{N,r}^{\rm succ}(|\alpha|^2)
&=
\frac{r^2}{|\alpha|^2}
\left[
\sum_{p,q:|p-q|=r}
|A^{(p,q)}|^2
\right]^N=
\frac{r^2}{|\alpha|^2}
\left[
\sum_{p,q:|p-q|=r}
e^{-|\alpha|^2}\frac{(|\alpha|^2/2)^{p+q}}{p!q!}
\right]^N =
\frac{r^2}{|\alpha|^2}
\Bigl[
2e^{-|\alpha|^2}I_r(|\alpha|^2)
\Bigr]^N.
\label{eq:PNr-exact}
\end{align}
Here $I_r(x)=\sum_{m=0}^{\infty}\frac{(x/2)^{2m+r}}{m!(m+r)!}$ is the modified Bessel function of the first kind. Since all accepted cases with $r\ge 1$ obtain the same normalized state after parity postselection, the total success probability is
\begin{align}
P_{N}^{\rm succ}(|\alpha|^2)
=
\sum_{r=1}^{\infty}
\frac{r^2}{|\alpha|^2}
\Bigl[
2e^{-|\alpha|^2}I_r(|\alpha|^2)
\Bigr]^N.
\label{eq:PN-exact}
\end{align}
\begin{table}[h]
\caption{Optimal coherent-state amplitude $|\alpha|_{\rm opt}^2$ and corresponding maximal conditional success probability $P_{N,\max}^{\rm succ}$ for the protocol of Eq.~\eqref{eq:PN-exact}. The optimization is only over the coherent-state amplitude $|\alpha|^2$ and is relative to the acceptance rule in Eq.~\eqref{eq:accepted-kraus}; it is not an optimization over all possible erasure or feedforward strategies. }
\label{tab:memory-success}
\begin{ruledtabular}
\begin{tabular}{ccc}
$N$ & $|\alpha|_{\rm opt}^2$ & $P_{N,\max}^{\rm succ}$ \\
\hline
2 & 0.773 & 0.219 \\
3 & 0.901 & 0.0764 \\
4 & 1.005 & 0.0303 \\
5 & 1.085 & 0.0126
\end{tabular}
\end{ruledtabular}
\end{table}

The maximal success probability of this protocol is therefore
\begin{align}
P_{N,\max}^{\rm succ}
=
\max_{|\alpha|^2>0}
P_{N}^{\rm succ}(|\alpha|^2).
\label{eq:PN-max-exact}
\end{align}
For $N=2,3,4,5$, the corresponding optimal values are listed in Table~\ref{tab:memory-success}.  The probabilities in Table~\ref{tab:memory-success} correspond only to the conditional success probability of the exact protocol considered here. They should therefore be viewed as illustrative of this specific construction rather than as optimized performance limits for memory-assisted telescopy.
\paragraph*{Measure.}
After the parity check, the memories encode a clean distributed logical qudit supported on $\mc H_m^{(1)}
= \mathrm{Span}\{\ket{1_j}_m\}_{j=1}^N$ with state proportional to Eq.~\eqref{eq:psi-after-parity}. Any desired rank-one projector $\ket{b}\bra{b}$ with
\begin{align}
\ket{b}=\sum_{j=1}^N b_j\ket{1_j}_m,
\end{align}
can then be implemented by a universal readout of this distributed-memory qudit, for example, by teleporting the memory state to a central node and measuring it there. We do not optimize the resources required for this final readout.


\end{document}